\documentclass{amsproc}
\usepackage[dvips]{graphicx}

\numberwithin{equation}{section}

\newcommand{\Figref}[1]{Fig.~\ref{#1}}
\newcommand{\Eqref}[1]{Eq.~(\ref{#1})}
\newcommand{\Secref}[1]{Section~\ref{#1}}

\begin{document}

\title{Spatiotemporal Chaos in Large Systems:\\
The Scaling of Complexity with Size}

\author{Henry S. Greenside}

\address{Department of Physics and the Center for
Nonlinear and Complex Systems, Duke University, Durham,
NC 27708-0305}

\email{hsg@phy.duke.edu}

\thanks{The author's research was supported in part by
NSF Grant DMS-9307893 and DOE Grant
DOE-DE-FG05-94ER25214.}

%    General info
\subjclass{Primary 54C40, 14E20; Secondary 46E25, 20C20}

\date{August 16, 1996}

\begin{abstract}

The dynamics of a nonequilibrium system can become
complex because the system has many components (e.g., a
human brain), because the system is strongly driven
from equilibrium (e.g., large Reynolds-number flows),
or because the system becomes large compared to certain
intrinsic length scales. Recent experimental and
theoretical work is reviewed that addresses this last
route to complexity. In the idealized case of a
sufficiently large, nontransient, homogeneous, and
chaotic system, the fractal dimension~$D$ becomes
proportional to the system's volume~$V$ which defines
the regime of extensive chaos. The extensivity of the
fractal dimension suggests a new way to characterize
correlations in high-dimensional systems in terms of an
intensive dimension correlation length~$\xi_\delta$.
Recent calculations at Duke University show
that~$\xi_\delta$ is a length scale smaller than and
independent of some commonly used measures of disorder
such as the two-point and mutual-information
correlation lengths. Identifying the basic length and
time scales of extensive chaos remains a central
problem whose solution will aid the theoretical and
experimental understanding of large nonequilibrium
systems.

\end{abstract}

\maketitle

\section{Introduction}
\label{intro}

Is the weather difficult to forecast because the world
is so big? Do human hearts, unlike those of mice,
sometimes spontaneously fibrillate because they are
large enough to sustain a different kind of dynamics
\cite{Winfree94}?  Are there limits to building an
arbitrarily powerful synchronous digital parallel
computer because a sufficiently large network of
coupled nonlinear computer clocks may become unstable
to asynchronous behavior?  These questions capture the
essence of some recent experimental and theoretical
efforts to understand how the dynamics of physical
systems may depend on their size. Progress in answering
these questions will likely be useful for engineering
applications in which one wants to design, control,
simulate, and forecast complex dynamical systems. But
progress in understanding these issues is also
rewarding in its own right. There is pleasure in
discovering unifying principles that can explain the
fascinating never-repeating patterns that are found in
many chemical, mechanical, biological, electronic,
fluid, and plasma driven dissipative systems
\cite{Cross93}.

In the following, I review some recent and ongoing
research, both experimental and theoretical, that is
trying to answer these kinds of questions concerning
the relation of dynamics to system size. For reasons
that are discussed further below, my review focuses on
a restricted class of sustained nonequilibrium systems,
viz., those that are nontransient, homogeneous, and not
driven strongly out of equilibrium. These conditions
represent idealizations and simplifications of the
complexities that occur in many natural and artificial
systems and so provide an important opportunity to
obtain insight about how one particular
mechanism---increase in size---may contribute to
overall complexity. A particular hope is that studying
the thermodynamic limit of infinitely large system
size, $L \to \infty$, may prove especially productive
for analyzing complex systems since various ideas from
statistical mechanics, thermodynamics, and
hydrodynamics may then apply
\cite{Hohenberg89,Cross93}. A related hope is that
subsystems of a finite inhomogeneous system, over
appropriate length and time scales, may be
quantitatively understood in terms of what one learns
from subsystems of infinite homogeneous systems. Future
research will need to address the challenging questions
of transient dynamics, spatial inhomogeneities, and
other routes to complexity that bring in more realistic
details.

To illustrate the kinds of insights that one might
obtain from a thermodynamic approach to large
nonequilibrium systems, we can suggest a partial answer
to the above question of whether the earth's weather is
hard to predict because the world is so big. We
idealize and assume that the earth's atmosphere can be
treated as a homogeneous turbulent medium. We then
carry out a thought experiment of letting the radius of
the earth, and so its surface area, increase
indefinitely while somehow keeping the composition of
the atmosphere and its physical processes unchanged.

In this limit and assuming homogeneity, thermodynamics
suggests that we think about two kinds of quantities
for characterizing the dynamics \cite{Callen85}: {\em
extensive} quantities analogous to the energy, entropy,
or mass of an equilibrium system, whose values grow in
proportion to the system volume~$V$; and {\em
intensive} quantities analogous to the pressure,
temperature, chemical potential, or mass density of an
equilibrium system, whose values are independent of
system size (for a sufficiently large size). Extensive
and intensive quantities arise from the locality of
physical interactions \cite[Section 2]{Landau80}. A
sufficiently large subsystem of an infinite system is
coupled only weakly to other subsystems, by an amount
of order the ratio of its surface area (which
represents the region of coupling) to its volume. Over
some time scale that increases with increasing
subsystem
size\footnote{\label{relaxation-time-footnote}Systems
often relax towards equilibrium by diffusive processes
so that a relaxation time would scale algebraically
as~$L^2$, where~$L$ is the system size.  For decay of
transients towards the attractor of a sustained
nonequilibrium system, the scaling of relaxation time
with system size is poorly understood. In some cases,
one finds algebraic, but slower than diffusive,
scaling~\cite{Greenside84pra,Cross84} while in some
other cases the relaxation time grows exponentially
with system
size~\cite{Shraiman86,Crutchfield88,Lai95prl}.}, the
weak coupling implies that the subsystem dynamics will
be approximately nontransient and uncorrelated with the
dynamics of other subsystems.  In the thermodynamic
limit, various parameters are extensive since they are
proportional to the number of independent subsystems
(which are statistically identical by the assumption of
homogeneity), or are intensive since they characterize
a particular subsystem.

As discussed below in \Secref{measures}, theory and
numerical experiments indicate that the fractal
dimension\footnote{The reader is assumed to be familiar
with the concept of a fractal dimension~$D$ of the
measure of a dynamical system's attractor \cite{Ott93}.
It is a nonnegative number that is 0, 1,
integer-valued, or real-valued for respectively fixed
points, periodic behavior (limit cycles), quasiperiodic
behavior (tori), and strange attractors. When
calculated from time series, the dimension is useful to
know since the ceiling of~$D$ provides an estimate of
the minimum number of degrees of freedom of {\em any}
mathematical model that can generate the particular
dynamics\cite{Abarbanel93}}~$D$ of a large homogeneous
turbulent system such as our idealized atmosphere will
be extensive, increasing in proportion to the surface
area of the earth for a fixed atmospheric depth. In
contrast, the largest positive Lyapunov
exponent\footnote{The largest Lyapunov
exponent~$\lambda_1$ is a real-valued measure of
instability, giving the average rate of exponential
separation of two infinitesimally close initial
conditions in phase space \cite{Ott93}. A positive
exponent~$\lambda_1 > 0$ is a commonly stated necessary
criterion for a deterministic dynamical system to be
chaotic.}~$\lambda_1$ is an intensive quantity,
becoming independent of system size~$L$ for large
enough~$L$.

One can then argue that short-term meteorological
forecasting is limited {\em not} by the large size of
the atmosphere but rather by its many interacting
components and their local nonlinear couplings. This
follows since the forecasting time~$1/\lambda_1 \approx
\mbox{2 weeks \cite{Lorenz93book}}$ is an intensive
quantity, independent of system size in the
thermodynamic limit.  Similar thermodynamic reasoning
suggests ways to test other potential relations between
quantities used to describe spatiotemporal dynamical
systems. Thus some researchers
\cite{Rasmussen87,vandeWater93} have speculated about
possible relations between the reciprocal of the
correlation time, $1/\tau$, of time series obtained
from a chaotic system and the largest Lyapunov
exponent~$\lambda_1$ or between~$1/\tau$ and the metric
entropy~$H$ (a dynamical quantity that quantifies
instability by the rate at which new information is
created \cite{Ott93}). But as pointed out by Sirovich
and Deane \cite[page 262]{Sirovich91}, the time~$\tau$
is an intensive quantity and so should be related to
other intensive quantities such as the
exponent~$\lambda_1$ or to the entropy density $h =
\lim_{V \to \infty} H/V$, rather than to extensive
quantities such as the entropy~$H$. Similarly, the
asymptotic exponential decay of the high-frequency part
of the power spectrum~$P(\omega)$ associated with a
bounded deterministic system
\cite{Frisch81,Greenside82} should have an intensive
decay rate. Verifying this intensive behavior could
resolve some conjectures by Sigeti \cite{Sigeti95},
that this decay rate may be related to the
intensive~$\lambda_1$ or to the extensive entropy~$H$.

The assumptions of homogeneity and extensivity invoked
in these arguments raise numerous interesting
questions: is there a characteristic length scale above
which a homogeneous nonequilibrium system becomes
extensively chaotic ($D \propto V$) and, if so, what
determines this length scale?  In particular, is the
earth's atmosphere extensively chaotic and
is~$\lambda_1$ indeed an intensive meteorological
quantity? As system size is increased (with all other
parameters held fixed), is there a gradual or abrupt
transition to extensive scaling? Are intensive
quantities such as~$\lambda_1$ {\em local} quantities
in that their values can be calculated from information
localized to some region of space\footnote{The
temperature~$T$ of an equilibrium system is intensive
as a partial derivative at constant volume,
$T=(\partial{E}/\partial{S})_V$, of extensive
energy~$E$ with respect to extensive entropy~$S$. But
the temperature can be measured locally, e.g., with a
thermometer, without having to know first the global
relation~$E=E(S,V)$.}? If so, by what algorithms?
Perhaps most importantly, what intensive quantities
should be used for characterizing large nonequilibrium
systems?  The correlation functions, fractal dimensions
and Lyapunov exponents discussed below in
\Secref{measures} are logical, rather than physical,
choices suggested by mathematical theory. They lack the
deeper significance of quantities such as energy,
entropy, and temperature which ultimately follow from
conservation laws of additive quantities
\cite{Landau80}.  Nonequilibrium systems generally lack
conservation laws because of the fluxes of energy and
matter required from external sources to drive systems
from equilibrium.

In the absence of a fundamental theory of sustained
nonequilibrium systems
\label{fundamental-theory-mentioned} that can suggest
the appropriate quantities to measure, this review will
consider only some simple-minded ways to characterize
spatiotemporal disorder, viz., several different kinds
of correlation lengths. Although such lengths are crude
ways to summarize the information available in
correlation functions, and correlation functions
themselves only represent lower-order statistical
information of some probability distribution function,
the study of correlation lengths is an important first
step that can indicate which basic length scales are
relevant in a large nonequilibrium system. A more
refined analysis may occasionally be justified by the
details of a particular experiment (e.g., the
statistics of defects in B\'enard convection
\cite{CrossMeironTu94,Ecke95}) but further progress
will ultimately depend on new theory.

As discussed below in \Secref{measures}, the extensive
scaling of the fractal dimension with system volume
suggests a relatively new measure of spatiotemporal
disorder, the dimension correlation length~$\xi_\delta$
introduced by Cross and Hohenberg \cite[Page 948,
Eq.~(7.38)]{Cross93}.  Researchers from my group at
Duke University, using a powerful parallel computer,
have provided some of the first systematic studies
of~$\xi_\delta$ for models of homogeneous
spatiotemporal chaos
\cite{Egolf94nature,Egolf95prl,OHern96,OHern97}.
These calculations yielded the somewhat unexpected
result that the length~$\xi_\delta$ is generally
independent from, and smaller than, the correlation
lengths derived from the more familiar two-point and
mutual information \cite{Fraser89} correlation
functions.

This result has implications for physics and for
simulation. One is that at least {\em two} length
scales are needed to characterize homogeneous
spatiotemporal chaos. Another is that the concept of
correlation is not so straightforward and care is
required to choose the appropriate definition in some
given context. A third implication is that the
length~$\xi_\delta$ suggests one answer to what is
meant by a ``big'' nonequilibrium system, namely one
for which the system size~$L$ satisfies~$L \gg
\xi_\delta$.  Fourth, it is natural to speculate that
the length~$\xi_\delta$ is the characteristic size of
weakly interacting subsystems.  If so, this has
implications for using parallel computers to simulate
the long-time dynamics of large spatiotemporal chaotic
systems since one could replace a long time integration
(which is difficult to parallelize) with a shorter time
integration over a much larger spatial region (which is
easier to parallelize using domain decomposition). By
analogy to ergodicity which allows one to replace
ensemble averages by time averages, this could be
called ``spatial ergodicity'' since the many
independent subsystems in the large domain would
constitute different realizations of the dynamics over
the length scale~$\xi_\delta$.

The rest of this review is divided into the following
sections. In \Secref{complexity}, several routes are
discussed that lead to increased dynamical complexity.
It is argued that the scaling of complexity with size
is a particularly promising route from a theoretical
point of view. In \Secref{expts}, some representative
large-aspect ratio experiments are discussed that have
guided our understanding of the thermodynamic limit of
spatiotemporal chaos. In \Secref{measures}, several
different correlation lengths are introduced, including
the dimension correlation length. The parametric
dependence of the different correlation lengths are
then discussed for numerical calculations on a
particular two-dimensional model of spatiotemporal
chaos. Finally, in \Secref{conclusions}, key points of
this review are collected and some questions identified
for further research.

This review is oriented towards the other participants
of this workshop, i.e., mainly towards fluid
dynamicists who have worked more in the area of
Navier-Stokes turbulence than in nonlinear dynamics and
spatiotemporal chaos. For this reason, I have tried to
be somewhat more pedagogical in motivating the
questions asked and ideas used. The books by Strogatz
\cite{Strogatz94} and Ott \cite{Ott93} are good
references for nonlinear dynamics of low-dimensional
systems while the book by Manneville
\cite{Manneville90} and the review article by Cross and
Hohenberg \cite{Cross93} provide good surveys of the
nonequilibrium physics of spatially extended
systems. Readers might also find useful a more detailed
knowledge of thermodynamics and statistical physics
than is usually discussed in fluid dynamics texts,
e.g., at the level of Landau and Lifshitz
\cite{Landau80} or of Callen \cite{Callen85}.

\section{Routes to Increased Complexity}
\label{complexity}

A specific motivation for investigating how dynamics
may scale with system size is the fact that it remains
extremely difficult to understand many of the complex
nonequilibrium systems in nature or even in the more
controlled settings of laboratory experiments such as
Rayleigh-B\'enard convection \cite{Busse78,Cross93}.  A
reasonable strategy might then be to study separately
some of the different mechanisms that increase the
complexity of nonequilibrium systems and then to study
what happens when these different mechanisms are
combined. (For now, I will avoid trying to pin down the
slippery concept of ``complexity'', which will be
discussed briefly in \Secref{measures}.) In this
section, three routes to increased complexity will be
discussed for sustained nonequilibrium systems:
increasing the number of interacting components,
driving the system further from equilibrium, and making
the system large with other parameters held fixed.

Before discussing these routes, it will be useful to
clarify what is meant by ``understanding'' the
spatiotemporal dynamics of a sustained nonequilibrium
system and why our understanding is presently so
poor. An essential prerequisite for understanding any
physical system is that one can define its state
mathematically and measure its state physically.  How
to define the state of a system is often {\em not}
clear and can be ambiguous, e.g., one can choose a
microscopic (molecular) or a macroscopic (continuum)
description of the same fluid. In either case, the
state of a system is a set of numbers which is
necessary and sufficient for solving certain known
dynamical equations. (The set of numbers and the
appropriate equations must often be discovered together
which historically has been difficult in many cases.)
Once the state of a system is known, one can ask
further questions such as: how to classify different
states, how states change with parameters, what kind of
bifurcations separate one state from another, how does
the stability of states change with parameters, and how
does transport of energy, momentum, or mass depend on a
state.

The difficulty of understanding spatiotemporal chaos
lies precisely in finding a sufficiently {\em reduced}
description of the system in terms of states that are
mathematically well-defined, computationally
accessible, and physically measurable. The
Navier-Stokes equations are a valuable reduction of
molecular dynamics but their turbulent solutions are,
in turn, so complex that a further reduction and
simplification would be useful at still longer
wavelengths and over longer time scales. Some tentative
initial steps have been taken in finding such
reductions\label{page:reductions}
\cite{Zaleski89,Bourzutschky92,Miller93,Chow95,Dankowicz96}
but our knowledge remains quite limited. Identifying
the appropriate length scales over which a further
reduction might be constructed, or even knowing whether
such length scales exist, is a central unsolved
problem.

Systems in thermodynamic equilibrium are perhaps an
outstanding example to keep in mind that are well
understood in the sense of knowing how to specify
states. Thermodynamic theory tells us that any {\em
two} macroscopic thermodynamic parameters, say
pressure~$p$ and temperature~$T$, suffice to
characterize an equilibrium system \cite{Landau80} and
remaining parameters can be determined from equations
of state.  For given thermodynamic parameters like~$p$
and~$T$, equilibrium states are not necessarily
homogeneous and can exist in different
phases\footnote{Recall that phases are distinct states
of the same substance that can be in equilibrium with
each other. Phases can be distinguished by additional
quantities called order parameters \cite{Landau80},
e.g., a mass density can distinguish solid, liquid, and
gas phases while a magnetization density can
distinguish paramagnetic and ferromagnetic phases of
iron.} and most substances exist in only a modest
number of phases such as solid, liquid or gas
phases. (Some substances such as metallic plutonium
have many crystalline phases but still no more than of
order one hundred.) Phases can undergo transitions as
parameters are varied and the stability of a phase can
be calculated from knowledge of its free energy.  The
phase transitions are mainly of two kinds, first-order
(subcritical or discontinuous) and second-order
(supercritical or continuous). The theory of critical
phenomena \cite{Binney92} provides a powerful
quantitative and experimentally tested method to
characterize second-order phase transitions.  There are
critical exponents that describe how order parameters
scale near a second-order transition, and these
exponents have universal properties in that they depend
only on a few details such as the lattice symmetry and
spatial dimensionality of the equilibrium system.

By comparison, our understanding of low-dimensional
nonequilibrium systems is already in a much less
satisfactory state, not even worrying about higher
dimensional, spatially extended systems which is the
concern of this review. While a ``microscopic'' state
for a given set of ordinary differential equations
(abbreviated as o.d.e.s in the following) is readily
defined in terms of a point in some phase space, it is
the ``macroscopic'' long-time description of the
phase-space attractors that is not well understood.  By
using Fourier analysis to compute the power spectrum of
a time series of some dynamical system, the
corresponding attractor can be classified simply and
efficiently as being stationary, periodic,
quasiperiodic, or nonperiodic \cite{Gollub80}. With
some effort\footnote{The further classification by
  Lyapunov exponents and by the Lyapunov fractal
  dimension~$D$ is computationally expensive for
  large~$D$ \cite{OHern96} and remains
  impractical except for known mathematical equations
  in one- or two-space dimensions and for researchers
  with access to powerful computers.}, the nonperiodic
states can be partially further classified, e.g., by
their Lyapunov spectrum or by their fractal dimension
\cite{Ott93}. At this time, a mathematical
characterization and classification of attractors and
of their bifurcations remains incomplete, especially of
bifurcations from a non-chaotic state to a chaotic
state or from one chaotic state to another. For
spatially extended high-dimensional systems, even less
is known about the kinds of attractors and bifurcations
that can occur \cite{Bunimovich91,CrossHohenberg94}. It
is an open and important question whether there are
macroscopic parameters analogous to temperature or
entropy that give a concise, useful, and computable
description of spatiotemporal chaos at long
wavelengths.

I have taken some time to discuss what are largely
elementary observations about the challenge of defining
or measuring states of nonequilibrium systems since
these are central issues in learning how to quantify
such systems.  The earth's atmosphere is a good example
of a nonequilibrium system that remains poorly
understood because we do not have useful ways to
characterize its complex dynamics. The following is a
partial list of why atmospheric dynamics is complex:
the atmosphere is made up of many physical components
such as nitrogen, carbon dioxide, water vapor, and
ozone, leading to many coupled spatiotemporal fields
such as temperature, pressure, velocity, and
concentrations; the atmosphere is inhomogeneous through
the presence of clouds and through coupling to
difficult-to-characterize topography such as deserts,
forests, ocean surfaces, and ice caps; the earth's
atmosphere has a large geometric aspect ratio (ratio of
lateral width to depth) and thus permits dynamics on
long wavelengths; and the atmosphere is strongly driven
out of equilibrium by heating from the sun and so is
highly turbulent, involving a huge range of length and
time scales. Also, the atmosphere remains difficult to
observe since it is immense compared to the regions of
earth where scientists live or can make measurements.

From this meteorological example, we can identify at
least three distinct routes to increased complexity
that (unfortunately for theorists) typically occur
together in many nonequilibrium systems. These routes
are: many interacting components, strong driving from
equilibrium, and large size. We discuss these briefly
in turn and then concern ourselves with this last route
for the remainder of this article.

\subsection{Complexity Arising From Many Different Components}
\label{many-components}

One route to complexity arises by coupling more and
more different components together. This kind of
complexity can occur whether or not there are spatially
dependent variables, e.g., many elements can be wired
together in an electronic circuit or all the chemicals
of a chemistry set can be stirred together to see what
happens. Adding more and more components together
corresponds mathematically to coupling more and more
o.d.e.s and studying dynamics in a higher- and
higher-dimensional phase space. The brain is complex in
this way because of its many different components:
there are many different kinds of neurons communicating
chemically via many different kinds of
neurotransmitters, and the neurons further communicate
electrically though action potentials whose information
content possibly depends on the geometric shapes of the
different neurons.

Presently, there is little useful knowledge about how
any measure of complexity scales with the number of
different components in a dynamical system, nor is it
even clear how to pose this problem in a productive
way. Three of the more famous unsolved problems in
science are linked to this approach to increased
complexity.  One is the origin of life, which possibly
arose spontaneously out of a prehistoric chemical soup
made up of sufficiently many reacting chemicals
\cite{Kauffman93}. A second problem is the
characterization and stability of biological ecologies,
which consist of many species (components) interacting
in some given environment \cite{Murray93}. A third
problem is how the computing ability of neural tissue
arose (or arises) from the coupling of many neurons.

The mathematical and computational challenges of
dealing with many interacting components are
sufficiently severe that many researchers try to avoid
this regime by studying systems with few
components. This is one reason why fluid dynamical
experiments such B\'enard convection and Taylor-Couette
flow continue to play such an important role in
nonequilibrium physics research \cite{Cross93}, since
there is already plenty of interesting dynamics to
explore with just one component, say water or a gas.

\subsection{Complexity Arising From Strong Driving}
\label{strong-driving}

A second route to complexity is that of a system being
driven strongly from equilibrium, which other speakers
at this Workshop have talked about. For fluid
dynamicists, this is the familiar and challenging
territory of high Reynolds-number flows, e.g., fluids
pushed through a pipe at high velocity. This route is
actually less general than adding more components or by
increasing the system size \cite{Cross93} since many
physical systems change their properties substantially
when driven too far from equilibrium. For example,
chemical gradients in reaction-diffusion systems can be
increased only so far before some component may start
to precipitate out. The pumping field of a laser can be
increased only so far before a laser may be ruined by
an electrical discharge. Increasing the temperature
gradient across a system can lead to phase changes or
to chemical decomposition.

If strong driving can be attained, then a common
consequence is the creation of spatiotemporal structure
on ever smaller length scales compared to the size of
the experimental apparatus itself. As an example, for
fluid flow characterized by a sufficiently large
Reynolds number~$R$, the Kolmogorov theory of
homogeneous turbulence tells us that there is a length
that scales as~$R^{-3/4}$, below which the dynamics is
damped out by dissipation and so can be ignored
\cite[Section 33]{Landau87}. More generally, Cross and
Hohenberg have speculated that the range of length
scales (and so the fractal dimension) will grow
algebraically as~$R^a$ for sufficiently large~$R$,
where~$R$ is the parameter driving the system from
equilibrium and~$a$ is some positive constant
\cite[Eq.(7.3), page 941]{Cross93}.

Since in a Fourier representation of fields, an
increasing number of Fourier components become active
and coupled as the Reynolds number increases, one might
consider the limit of strong driving as a particular
case of the previously discussed route involving many
coupled components. But the situation here is different
since the addition and interaction of new modes with
increasing~$R$ corresponds to a parameter change of
fixed known equations (the Navier-Stokes equations)
while the increase in the number of components of a
chemical or ecological system corresponds to changing
the mathematical equations themselves.

\subsection{Complexity Arising From Large System Size}
\label{large-size}

The third route to complexity is perhaps the most
widely occurring and arises from making a system large
compared to certain natural length scales while keeping
fixed the number of components and the magnitudes of
thermodynamic gradients (which sustain the system in a
nonequilibrium state). This approach has been studied
quantitatively in the laboratory and by theorists only
relatively recently because only in the last decade or
so have computers become sufficiently powerful,
inexpensive, and widely available so as to allow
careful control of large experimental systems, and the
accumulation and analysis of large amounts of
spatiotemporal data.

What is meant by a ``large'' experimental system here?
For many experimental systems that are driven slowly
out of equilibrium by increasing some thermodynamic
gradient, there is a stable uniform state that becomes
linearly unstable to a nonuniform spatiotemporal state
with a cellular structure\footnote{Some hopefully
familiar examples are the appearance of convection
rolls from a motionless conducting fluid in a
Rayleigh-B\'enard experiment as a temperature
difference is increased, or the appearance of Taylor
vortices in Taylor-Couette flow from a featureless flow
as the angular frequency of the inner cylinder is
increased. Many other examples are given in the review
article of Cross and Hohenberg \cite{Cross93}.}. With
increased driving, the primary linear instability
typically occurs first at a critical
wavelength~$\lambda_c$ so that one can initially say
that an experimental system of size~$L$ is large when
its aspect ratio~$\Gamma = L / \lambda_c \gg 1$. In
\Secref{measures}, we will come across other larger
lengths (correlation lengths), compared to which the
system should also be large.

Presently, experimentalists can attain aspect ratios of
order~$10^2$ to~$10^3$ while maintaining fairly uniform
boundary conditions\footnote{In the same range of
magnitude is the aspect ratio~$\Gamma \approx 600$ of
the earth's troposphere (where weather occurs), as
estimated by the ratio of the earth's radius, $6400 \,
\rm km$, to a typical depth of the troposphere,
about~$10 \, \rm km$ \cite{Morrison95}).}
\cite{Cross93}. For certain model equations and using
gigaflop-class computers, computational scientists can
simulate one-dimensional aspect ratios of order~$10^4$
\cite{Sneppen92,Egolf94thesis} and two-dimensional
aspect ratios of order~$10^3$ \cite{Chate96-2dcgl}. As
impressive as these achievements are, these aspect
ratios are tiny compared to the effective aspect ratio
of order~$10^8$ for a macroscopic equilibrium crystal,
which may be one centimeter on a side compared to a
atomic lattice spacing of a few Angstroms. This
observation raises concerns whether experiments on even
the largest nonequilibrium systems are attaining
dynamics that are independent of the size and shape of
the experimental apparatus, or that are independent of
the choice of boundary conditions such as conducting or
thermally insulating lateral walls. Some of the results
discussed in \Secref{measures} concerning the dimension
correlation length suggest that aspect ratios
of~$\Gamma\ge 100$ are, in fact, sufficiently large for
the dynamics to be considered statistically homogeneous
and independent of boundary conditions but further
quantitative studies are needed.

A serious difficulty in studying the large-aspect-ratio
limit, either experimentally or computationally, is the
diverging time scale for transients to decay with
increasing system size. Thus in recent
large-aspect-ratio reaction-diffusion experiments of
Ouyang and Swinney concerning the onset of chemical
turbulence \cite{Ouyang91}, typical experimental
observation times were about a week or less. Yet the
lateral size of their reaction cell, $L = 2.5 \, \rm
cm$, combined with the typical diffusion constant of
their reagents, $D \approx 10^{-5} \, \rm cm^2/sec$,
gives a lateral chemical diffusion time~$\tau = L^2/D
\approx 1 \, \rm week$ and so their experiments are
only on the edge of being nontransient. Similarly,
lateral thermal diffusion times in convection
experiments can be of order days for large cells in
which case weeks or more of observation may be required
to establish that transients have sufficiently
decayed. Finding experimental systems with fast time
scales is then an important goal. Two leading
candidates from fluid dynamics are crispation
experiments (parametrically forced surface waves
excited by shaking a container of fluid up and down
periodically\footnote{Some pictures of spatiotemporal
dynamics in crispation experiments are available at the
web sites
\texttt{http://www.haverford.edu/physics-astro/Gollub/faraday.html}
and \texttt{http://cnls-www.lanl.gov/nbt/movies.html}.}
\cite{Gollub90}) and convection experiments of a fluid
near its critical point, which allows for extremely
thin layers of fluid to be studied without creating
substantial non-Boussinesq effects
\cite{Assenheimer93}. Electronic and laser systems can
be found with faster time scales but so far are less
useful for quantitative comparisons with theory since
they lack a mathematical description as accurate or as
well parameterized as the Navier-Stokes equations.

In addition to the long time scales for transients to
decay in large-aspect-ratio nonequilibrium systems, a
related complication is the lack of good diagnostics to
identify the presence or end of transient behavior.  In
some calculations \cite{Crutchfield88}, the transients
seem statistically stationary over long periods of time
and then change abruptly to a nontransient regime. As
mentioned in the footnote on
page~\pageref{relaxation-time-footnote}, experiments,
simulations, and theory suggest that the slowest time
scale for transients to decay in a nonequilibrium
system is sometimes not set by diffusion over the
largest lateral dimension~$L$, but by even slower
mechanisms that are not well understood.  This
slower-than-diffusive scaling is possibly a consequence
of intricately winding basins of attraction in a
high-dimensional phase space although a quantitative
demonstration of this idea is so far lacking.

\section{Large-Aspect-Ratio Experiments of
Spatiotemporal Chaos}
\label{expts}

In this section, a few experimental results are
summarized to indicate the kinds of insights that
experiments have provided about large nonequilibrium
systems, especially from convection experiments. Only a
few highlights will be given here since Cross and
Hohenberg have given an especially complete and
detailed review of convection experiments (as well as
of other experimental systems), to which the reader
should turn for more details and references
\cite{Cross93}. A key conclusion will be that
spatiotemporal chaos is a common phenomenon in
large-aspect-ratio nonequilibrium systems, even for
weak driving in the sense of maintaining a system just
above the onset of a primary bifurcation from a uniform
to non-uniform state. Visually, the time evolution of
various fields involves the creation, annihilation,
motion, and interaction of defects, which are regions
where a local spatial periodicity can not be defined.

Once a goal of studying the scaling of complexity with
size has been identified, the discussion of
\Secref{complexity} suggests that one should try to
find experimental systems that avoid the other routes
to complexity. This can be done by choosing systems
with as few components as possible (to reduce the total
number of coupled equations), by choosing systems that
are driven out of equilibrium as weakly as possible (so
that nonlinearities are weak and more easily treated
mathematically, and to reduce the range of length and
time scales that need to be considered), and by
choosing systems for which one can construct
high-precision homogeneous large-aspect-ratio
experiments with good visualization and good
reproducibility. These are difficult goals and it is
impressive that experimentalists have succeeded in
approaching these idealizations for a variety of
systems.

Theorists would add another goal for experimentalists,
which is to simplify the mathematical effort by
identifying systems that can be quantitatively
described by models with only one spatial
variable. This has turned out to be more challenging to
achieve than many researchers expected since boundary
conditions that restrict an experimental medium to
one-dimensional behavior often raise the threshold for
the onset of chaos to strong levels of
driving\footnote{An important exception is convection
of binary fluids in a large narrow annular domain
\cite{Cross93}. The addition of an extra component, a
concentration field, changes the onset of convection
dramatically to an instability with a nonzero imaginary
part for the eigenvalue with the largest real part,
leading to interesting one-dimensional dynamics near
onset involving traveling waves. Unfortunately, the
instability is also now weakly subcritical which makes
a quantitative theory more difficult and, for current
experiments with small Lewis numbers, the concentration
field introduces even longer time scales for transients
to decay.}.  Because of the challenges of finding
one-dimensional large-aspect-ratio systems with
interesting dynamics near onset, many experiments
continue to be done in $2 \frac{1}{2}$ dimensions, in
which there are two large lateral directions and a
short transverse direction. Few experiments or
simulations have been done in truly three-dimensional
large-aspect-ratio systems, although there is growing
interest in this topic, e.g., for researchers studying
excitable media who would like to know whether the
onset of fibrillation depends on the three-dimensional
structure of heart muscle \cite{Winfree95}.

\begin{figure}[ht] 
  \centering
  \includegraphics[width=3.5in]{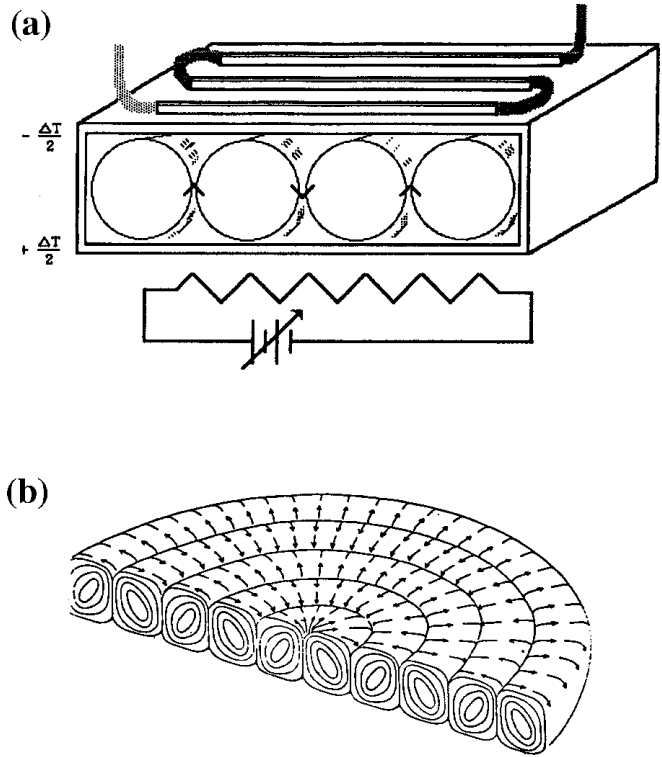}
\caption{ {\bf (a)} A schematic picture of a
Rayleigh-B\'enard convection apparatus. Two horizontal
flat plates of high thermal conductivity (typically
made of copper or sapphire) are separated by a vertical
height~$d$ and the gap is filled with a fluid such as
water. A spatially uniform and time-independent
vertical temperature gradient is imposed on the fluid
by heating the bottom plate to a fixed
temperature~$T_{\rm lower} = T+\triangle{T}/2$ (e.g.,
by a variable current flowing through resistors) and
cooling the upper plate to a temperature~$T_{\rm upper}
= T - \triangle{T}/2$ where~$T$ is the mean temperature
of the fluid.  The positive temperature
difference~$\triangle{T}$ across the fluid is the
bifurcation parameter for driving the system from
equilibrium. {\bf (b)} Cross-section of curved
convection rolls. For light directed vertically into
the fluid and reflected off the bottom plate, the
alternating warm and cold regions of convecting fluid
act as diverging and converging lenses, allowing the
spatial pattern to be visualized by shadowgraphy as
alternating light and dark bands (see
\Figref{fig:convection-patterns}). }
\label{fig:convection-apparatus} 
\end{figure}

Somewhat by trial and error and for historical reasons,
Rayleigh-B\'enard convection
(\Figref{fig:convection-apparatus})
%%%%%%%%%%
%%%%%%%%%%
has become a central paradigm for experimental and
theoretical research in large-aspect-ratio
nonequilibrium physics.  Convection does not quite
involve the minimum number of components possible in a
fluid experiment since there is a temperature field in
addition to the usual velocity and pressure fields of
the Navier-Stokes equations. But the presence of the
temperature field has the important consequence of
allowing a continuous bifurcation, from a uniform
motionless conducting state to a state of
finite-amplitude convection rolls in which warm fluid
rises and cold fluid falls spatially over a
characteristic wavelength that is basically twice the
depth~$d$ of the fluid, $\lambda_c \approx 2d$. Some
other experimental advantages of Rayleigh-B\'enard
convection are the ease of building highly precise
containers with static homogeneous boundaries, the ease
of visualization throughout the interior of the fluid,
the slow increase in complexity of the dynamics as the
temperature gradient is increased, and the absence of a
mean flow. As pointed out above, a drawback is the long
time scale for transients to decay, a lower bound for
which is given by the horizontal (lateral) thermal
diffusion time.

A convection experiment can be described by a minimum
of three dimensionless parameters \cite{Busse78}. The
most important is the Rayleigh number~$R = \alpha g d^3
\triangle{T} / \nu \kappa$ where~$\alpha$ is the
thermal-expansion coefficient of the fluid, $g$~is the
acceleration of gravity (the source of buoyancy
forces), $d$ is the depth of the fluid, $\triangle{T}$
is the vertical temperature difference across the
plates, $\nu$ is the kinematic viscosity, and~$\kappa$
is the thermal diffusivity. In a typical experiment or
simulation, all of the parameters entering the Rayleigh
number are fixed except the temperature
difference~$\triangle{T}$, and so the Rayleigh number
is conveniently thought of as being proportional to the
temperature difference across the plates. A linear
stability analysis of the conducting fluid in a
laterally infinite layer shows that convection begins
when the Rayleigh number increases beyond a critical
value~$R_c = 1707.76\cdots$, which corresponds to a
temperature difference of a few degrees Celsius in
room-temperature water experiments with fluid depths of
a few millimeters. Since the bifurcation to convection
is continuous, it is convenient to introduce a reduced
Rayleigh number, $\epsilon = (R -
R_c)/R_c$\label{eps-defn} which is a small parameter
near onset. Perturbative expansions in~$\epsilon$ near
onset lead to the amplitude equation formalism
\cite{Cross93} which has made many successful
quantitative predictions concerning pattern formation.

A second dimensionless parameter is the Prandtl
number~$\sigma = \nu/\kappa$, which is the ratio of the
strengths of the two channels of dissipation. The
Prandtl number influences the types of secondary
bifurcations which convection rolls undergo at higher
Rayleigh number, but weakly affects the dynamics
sufficiently close to onset. It is a fluid property
that is weakly dependent on temperature and is of order
one for many gases. Different experimental values can
be selected by using different fluids such as
large-$\sigma$ oils, moderate-$\sigma$ gases, and
low-$\sigma$ liquid mercury. A third dimensionless
parameter is the aspect ratio~$\Gamma = L / d$
(where~$L$ is the largest lateral distance), which
determines how weakly the lateral boundaries influence
the dynamics. For this review, the aspect ratio is the
interesting parameter since it provides a way to
increase dynamical complexity. Computational scientists
can simulate three-dimensional convection with periodic
lateral boundary conditions in which case only the
parameters~$R$, $\sigma$, $\Gamma$, together with
initial data, need to be specified. For
experimentalists who require real walls to contain a
fluid, other parameters must be considered such as the
thermal boundary conditions on the lateral walls (which
often have a thermal conductivity comparable to the
fluid) and the shape of the container. Experimentally
and theoretically, the influence of these other
parameters on dynamics is not well understood although
some experiments show that the choice of lateral
boundary conditions can strongly influence the
spatiotemporal dynamics \cite{Meyer91}. How this
influence varies with increasing aspect ratio is a key
question to resolve with future research.

The quantitative study of large-aspect-ratio dynamics
began in~1978, with a pioneering Rayleigh-B\'enard
convection experiment by Guenter Ahlers and Robert
Behringer who were then at AT\&T Bell Laboratories
\cite{Ahlers78ptps}. These researchers studied the
onset of convection in liquid helium at cryogenic
temperatures of a few degrees Kelvin\footnote{The
thermal conductivity of copper plates becomes large
compared to that of helium at such low temperatures so
that the assumption of a spatially-uniform
time-independent temperature gradient is especially
well satisfied.}. Using convection cells with three
different aspect ratios of about~2, 5,
and~57\footnote{The aspect ratio of~57 was an order of
magnitude larger than any previous experiment at the
time and was much larger than what any supercomputer of
that era could simulate.}, they made the remarkable and
unexpected discovery that the Rayleigh
number~$R_t=R_t(\Gamma)$ at which turbulent or chaotic
behavior was first observed (as the Rayleigh number was
increased in small constant steps from below to above
the onset of convection) decreased with
increasing~$\Gamma$, and seemed to approach the onset
of convection itself, $R_t \to R_c$, as the aspect
ratio became large. In fact, within experimental error
the Ahlers-Behringer experiment for~$\Gamma=57$ said
that the onset of chaos coincided with the onset of
convection, $R_t \approx R_c$. This last point was a
particularly exciting development since it showed for
the first time that a turbulent state existed in a
parameter range that might be quantitatively amenable
to perturbation theory (in the small parameter
$\epsilon = (R - R_c)/R_c \to 0^+$).

The discovery of chaos close to the onset of convection
in a large aspect-ratio cell was a big surprise at the
time and is still not well understood today, eighteen
years later. The reason for the surprise was that a
linear stability theory of periodic two-dimensional
roll solutions of the Boussinesq equations in a
laterally infinite container, had been developed over
many years by Fritz Busse and collaborators
\cite{Busse78}. This theory predicted that stable
time-independent periodic convection rolls should exist
sufficiently close to onset, with the roll wavelengths
lying in a known range that depends on both~$R$
and~$\sigma$ \cite{Schluter65}. This prediction had
been qualitatively confirmed by prior experiments in
smaller aspect ratio cells and was therefore expected
to be more quantitatively correct in the limit of large
aspect ratio, which approached the theoretical
assumption of a laterally infinite container. Busse's
theory further predicted that, with increasing~$R$,
{\em all} periodic two-dimensional rolls eventually
became unstable above some Rayleigh
number~$R_u=R_u(\sigma)$ although some rolls may first
become unstable for values of~$R$ smaller than~$R_u$
depending on their wavelength.  What happens for $R >
R_u$ (all two-dimensional periodic rolls unstable) is a
long-time nonlinear effect not predicted by a linear
analysis: the unstable rolls could evolve into a new
stationary pattern (e.g., consisting of curved
convection rolls) or into a sustained time-dependent
state. In any case, the linear stability theory
together with prior experiments suggested that any
time-dependent behavior, and certainly any chaotic
behavior, would not be seen until the reduced Rayleigh
number~$\epsilon = (R - R_c)/R_c$ was of order one, and
this result should hold especially in the limit of
large aspect ratio.

\begin{figure}[ht] 
  \centering
  \includegraphics[width=3.5in]{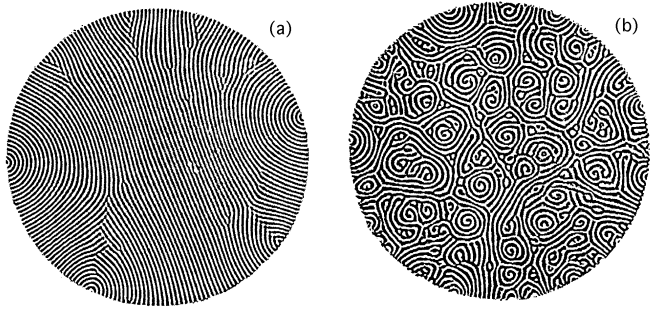}
\caption{Two instantaneous Rayleigh-B\'enard convection
patterns from different spatiotemporal chaotic states,
in a~$\Gamma = 74.6$ cylindrical convection cell
\cite{Morris96}.  The patterns were observed by
shadowgraphy; the white areas correspond to cool gas
descending into the plane of the figure and the dark
areas correspond to warm gas rising out of the plane of
the figure. Pattern {\bf (a)} corresponds to a fixed
reduced Rayleigh number~$\epsilon = 0.243$ while
pattern~{\bf (b)}, a spiral-defect chaos state, is more
strongly driven with~$\epsilon = 0.894$. These chaotic
states occur in a range of reduced Rayleigh
number~$\epsilon < \epsilon_u \approx 2$ for which
theory predicts the existence of stable periodic
infinite two-dimensional rolls.  The convecting fluid
is gaseous carbon dioxide at a mean temperature of
about~$30^\circ \, \rm C$, with Prandtl number~$\sigma
= 0.94$, and under a pressure of~$p = 33 \, \rm
bar$. The local periodicity of the convection rolls,
disrupted by defects, is a general feature of many
large spatiotemporal chaotic states.}
\label{fig:convection-patterns}
\end{figure}

An irony of this pioneering experiment was that no
visual information was available to indicate the
spatial patterns of the convecting fluid.  Ahlers and
Behringer had obtained their conclusions by analyzing
time series of the Nusselt number, which is a
dimensionless scalar measure of the total heat
transported from the bottom plate to the top plate of
the convection experiment. Not until 1985, with an
innovative high-pressure gas experiment of Pocheau,
Croquette, and Le Gal \cite{Pocheau85}, did researchers
succeed for the first time in visualizing
large-aspect-ratio convection patterns. The details of
that experiment largely exceeded the imagination of
those who had tried to guess the mechanism of the time
dependence in the Ahlers-Behringer experiment.  Some of
these details are illustrated in
\Figref{fig:convection-patterns}, which is taken from a
recent and more refined version of a pressurized gas
convection experiment\footnote{The time evolution of
these chaotic states can not be easily appreciated from
the snapshots in \Figref{fig:convection-patterns} but
fortunately there are animations of these and other
spatiotemporal chaotic states available via the World
Wide Web (WWW) at the following addresses::
\begin{center}
\begin{tabular}{l}
\texttt{http://www.cs.duke.edu/research/visualization.html}  \\
\texttt{http://www.cco.caltech.edu/\~{}mcc/st\_chaos.html}\\
\texttt{http://mobydick.physics.utoronto.ca/gallery.html}
\end{tabular}
\end{center}
To view these animations, a WWW~viewer such as Netscape
is needed which is configured for playing MPEG files
(e.g., via the free utility \texttt{mpeg\_play}).}
\cite{Morris96}.
%%%%%%%%%%%%%%%%%%%%%%%%%%%%%%%%%%%%%%%%%%%%%%%%%%%%%%%%%%%%%%%%%
%%%%%%%%%%%%%%%%%%%%%%%%%%%%%%%%%%%%%%%%%%%%%%%%%%%%%%%%%%%%%%%%%
Patterns (a) and~(b) immediately suggest (as did the
earlier experiment of Pocheau et al) why the Busse
stability analysis did not apply to the
Ahlers-Behringer experiment, although both for
different reasons.  \Figref{fig:convection-patterns}(a)
shows that even for this large aspect ratio, the
lateral boundaries can have a strong effect on the
geometry of the convection rolls, which have a
substantial curvature and so are three-dimensional over
much of the interior of the convection cell.  This
curvature can be understood qualitatively as a
consequence of the cylindrical geometry, of the
empirical fact that convection rolls tend to be
perpendicular to the side walls for certain lateral
thermal boundary conditions, and of the fact that near
onset, the rolls oscillate spatially within a narrow
range of wavelengths centered on the critical
wavelength~$\lambda_c \approx 2d$. Further from onset,
but still below the Rayleigh number for which all
two-dimensional infinite periodic rolls are unstable,
\Figref{fig:convection-patterns}(b) shows that the
Busse stability analysis can fail because the basic
pattern can be so disordered that there are no
substantial regions of straight rolls.  Measurements
suggest that the chaotic state in the interior of
\Figref{fig:convection-patterns}(b) does not depend on
the shape or size of the convection cell
\cite{Morris96} and thus this state (unlike
\Figref{fig:convection-patterns}(a)) is a good
candidate for studying the thermodynamic limit of
infinite system size. Despite the intricate
time-dependent structure of both patterns, they still
consist of locally periodic rolls, with a distribution
of local wavelengths that is almost entirely stable
according to Busse's linear stability analysis
\cite{Heutmaker85,Morris93}.

Since the above~1978 and~1985 experiments, many other
experiments, simulations, and analytical
calculations---on convection and on other
systems---have shown that large homogeneous sustained
nonequilibrium systems with spatiotemporal chaos have
various features in common that we briefly summarize
\cite{Cross93}:
\begin{enumerate}

\item
Sufficiently large systems {\em typically} have chaotic
solutions and the chaos can occur close to the primary
instability from a uniform state as the system size
increases. Bigger systems seem to be more susceptible
to chaos although a precise statement of this fact has
yet to be made.

\item
The spatiotemporal dynamics is often associated with
defects (regions where a local periodicity can not be
defined \cite[Section V.B, page 898]{Cross93}), whose
dynamics seem to determine both the spatial and
temporal disorder (but {\em not} the fractal dimension
as discussed in \Secref{measures}
\cite{OHern96}). These defects are evident in
\Figref{fig:convection-patterns}(a), which has focus
singularities, dislocations, disclinations and grain
boundaries, and in \Figref{fig:convection-patterns}(b)
which has in addition spiral and target defects. The
time evolutions of these states show that the defects
can nucleate, annihilate, move, and interact with other
defects and with lateral boundaries. Over the past
fifteen years, researchers have studied the dynamics of
isolated defects and of isolated pairs of defects, with
the hope (not yet realized) that insights of few-defect
dynamics might apply to many-defect states such as
\Figref{fig:convection-patterns}(b). A good example of
this approach is the recent theoretical paper by Cross
and Tu \cite{Cross95} on the spiral-defect chaos state
of \Figref{fig:convection-patterns}(b).

\item
From both experimental and computational studies of
large-aspect-ratio systems, there seem to be only a
{\em few} bifurcations that lead from the uniform state
to spatiotemporal chaos as the system is driven further
from equilibrium\footnote{An unusually thorough mapping
of nonequilibrium phases, and of the bifurcations
separating these phases, has been experimentally
determined by Andereck et al for Taylor-Couette flow
with both inner and outer cylinders rotating, for
aspect ratios ranging from~20 to~48 (see Fig.~1 of
Ref.~\cite{Andereck86}. It would be interesting to
determine whether their phase boundaries correspond to
those of the thermodynamic limit~$\Gamma \to
\infty$.}. After the onset of spatiotemporal chaos,
further distinguishable bifurcations are rarely
observed, although this may reflect our present
inability to identify such bifurcations
theoretically. These qualitative observations suggest,
somewhat paradoxically, that the thermodynamic limit
may be easier to understand than low-dimensional chaos
since there exist only a few phases in that limit
\cite{Shraiman92}. For small-aspect-ratio
low-dimensional models, there can be an infinity of
bifurcations over a finite parameter interval, e.g.,
for the logistic map or for the Lorenz equations
\cite{Strogatz94}.

\item There have been few examples discovered or
  studied in which nontrivial critical exponents occur
  at a continuous bifurcation in a large-aspect-ratio
  system. One example is the bifurcation of a chaotic
  to uniform (laminar) state via spatiotemporal
  intermittency, which was predicted to lie in the
  universality class of directed percolation \cite[Page
  954]{Cross93}. Experiments and simulations did not
  confirm this prediction although some details need to
  be studied further. A more recent example (discussed
  in \Secref{measures} in another context) is a study
  by Miller and Huse of a two-dimensional deterministic
  spatiotemporal chaotic nonequilibrium lattice that
  undergoes an Ising-like phase transition as a
  coupling constant is varied \cite{Miller93}. These
  authors used analytical and numerical calculations to
  argue that this deterministic nonequilibrium model
  should behave exactly like the equilibrium
  two-dimensional Ising model at long wavelengths
  (i.e., near the Ising phase transition). Some doubt
  was recently cast on this conclusion by ambitious
  numerical simulations of the Miller-Huse model by
  Marcq et al \cite{Marcq96}, who found evidence
  for critical exponents that are close to, but
  distinctly different than the Ising values.  This
  discrepancy is important to resolve since it would
  help to identify when equilibrium statistical
  mechanics might apply to sustained nonequilibrium
  systems. A related area of future research would be
  to understand the possible relevance of probabilistic
  (as opposed to deterministic) nonequilibrium models
  for understanding deterministic dynamics. There has
  been considerable research concerning the critical
  properties of probabilistic models, which show a
  wider variety of universality classes than those
  found in equilibrium systems \cite{Schmittmann95}.

\end{enumerate}

We finish this section by coming back to one of the
questions raised by the Ahlers and Behringer
experiment: is it true that the onset of chaos
approaches the onset of convection in the limit of
infinite aspect ratio: $R_t \to R_c$ as~$\Gamma \to
\infty$? This seems to be a delicate question that will
be difficult to settle by experiment or by simulations,
because of the diverging transient time as~$\Gamma$
becomes large. To some extent, interest in this
question has been preempted by more recent theory and
new experiments which show that, for a convection cell
rotating at a constant angular frequency~$\Omega$, the
onset of chaos should rigorously coincide with the
onset of convection above a critical rotation
speed~$\Omega_c$. This system thus seems to be
particularly attractive for studying the thermodynamic
limit since analytical progress should be possible near
onset \cite{CrossMeironTu94}.

\section{Characteristic Length Scales of Spatiotemporal Chaos}
\label{measures}

From the discussion of the previous section, we see
that a key challenge for understanding how complexity
scales with size, for large homogeneous sustained
nonequilibrium systems, is to find ways to characterize
the spatiotemporal disorder that is commonly observed.
This problem is difficult physically and
mathematically. The lack of conservation laws, detailed
balance, and free-energy-like functionals has hindered
the development of a basic theory of sustained
nonequilibrium systems that could indicate appropriate
quantities to measure, e.g., ones perhaps analogous to
scalar macroscopic equilibrium parameters such as the
temperature or chemical potential.  The nonperiodicity
of spatiotemporal chaos makes the problem of
characterization mathematically open-ended since there
is an infinity of ways to characterize slippery
concepts like ``disordered'', ``random'' \cite[Section
3.5]{Knuth81} and ``complex'' \cite{Zurek90}. For
example, random fields can be described incompletely by
an infinity of~$n$-point correlation functions (of
which only the~$2$-point function is routinely
computed), and a high-dimensional strange attractor
(perhaps the source of the random fields) can be
described incompletely by the infinity of multifractal
dimensions~$D_q$ \cite{Ott93} (of which only one value,
the correlation dimension $D_2$, is routinely
calculated). Further, many proposed mathematical
definitions for ``random'' are not computable from
finite amounts of imprecise data and, even when they
are, they have not been compared with each other in a
systematic statistical way. Given these difficulties,
it is better to look carefully at specific experimental
and computational results such as
\Figref{fig:convection-patterns} for guidance in posing
theoretical questions. As an example,
\Figref{fig:convection-patterns}(b) shows spirals of a
certain characteristic average width and this width
turns out to vary with Rayleigh number~$R$ and with
Prandtl number~$\sigma$. For this particular
experiment, one could then try to predict how this
width varies with parameters \cite{Cross95}.

In this section, this more modest approach is pursued
and we consider some recent efforts to answer a basic
question suggested by experiments and simulations: what
length scales are associated with homogeneous
spatiotemporal chaos?  If various length scales are
identified, one can further ask how they vary with
parameters and with each other. The calculations
discussed below show that there are at least two
independent length scales associated with homogeneous
spatiotemporal chaos. One length scale, the two-point
correlation length~$\xi_2$, is determined by the
time-averaged spatial disorder as measured by the
two-point (or the mutual information
\cite{OHern96}) correlation function of a
field. The second length scale, the dimension
correlation length~$\xi_\delta$, arises from the
extensivity of fractal dimensions in phase space.  We
discuss these two length scales in turn and then
demonstrate that they are independent by studying a
nonequilibrium model for which the length~$\xi_2$
diverges while the length~$\xi_\delta$ remains finite
as a parameter is varied \cite{OHern96}. Some
implications of the fact that these length scales are
independent from one another are then discussed.

\subsection{What does a two-point correlation length~$\xi_2$ mean?}
\label{meaning-of-xi_2}

The two-point correlation function of a random field
has a long tradition of application in fluid dynamics
and in condensed matter physics because of the ease
with which it can be computed and because theory is
sometimes able to predict some of its
features. However, its meaning in a spatiotemporal
context is not so clear as we now discuss. To introduce
some notation and definitions, let us consider a
nonperiodic spatiotemporal scalar field~$u(t,x)$ which
is assumed for simplicity to be one-dimensional with
zero mean.  (The field~$u(t,x)$ could be the
temperature field fluctuations~$T(t,x,y_0,z_0) -
\langle T\rangle$ measured along some~$x$-coordinate in
a B\'enard convection cell.) For spatial
coordinates~$x'$ and~$x$, the two-point correlation
function~$C_2(x',x)$ is defined by
\begin{equation}
  C_2(x',x) = \left\langle u(t,x') u(t,x) \right\rangle
  , \label{2-pnt-fn}
\end{equation}
where the brackets~$\langle \cdots \rangle$ denote
averaging over the time variable~$t$ (and perhaps also
over realizations calculated independently on different
processors of a parallel computer
\cite{Egolf94thesis}).  For the homogeneous systems
that we assume, $C_2=C_2(|x'-x|)$ is a function of only
the magnitude of the difference between the spatial
positions. In many (but not all) cases, the two-point
correlation function asymptotically decays
exponentially for a sufficiently large spatial
separation,
\begin{equation}
  C_2( |x' - x| ) \to \exp(- |x' - x|/\xi_2 ) , \qquad
    \mbox{as $|x'-x| \to \infty$} . \label{xi2-defn}
\end{equation}
which defines the two-point correlation
length~$\xi_2$. Fig.~4 of Ref.~\cite{OHern96}
illustrates this exponential decay for the
two-dimensional Miller-Huse model \cite{Miller93}, from
which the length~$\xi_2$ can be extracted from a
log-linear plot.

An important point to appreciate in the definition
\Eqref{2-pnt-fn} is that the definition of~$C_2$ is
only weakly dependent on time correlations of spatial
fields.  If one considers \Eqref{2-pnt-fn} as averaging
over various snapshots at different times~$t$ of the
product~$u(t,x') u(t,x)$, then the {\em order} of the
snapshots does not matter in the summation. It is then
unclear how the spatial disorder measured by~$\xi_2$
relates to the deterministic dynamics that generates
the disorder. As an example, meteorologists since the
1950's have measured two-point correlation functions of
the earth's weather patterns, e.g., of the varying
heights on a constant pressure surface \cite[Figure 5,
page 507]{Buell58}. (The author is indebted to Edward
Lorenz for this reference.) The correlations measured
from a specific point do not fall off exponentially
rapidly with increasing distance\footnote{Nor do
  meteorological correlation functions decrease
  substantially over the entire earth, which leads to
  the concept of teleconnections between different
  points on the surface of the earth \cite{Read93}.}
but do fall off substantially over a length scale of
about~$1500 \, \rm km$, which we take as an estimate of
the two-point correlation length~$\xi_2$.  If the
weather at two different points is {\em dynamically}
(not just spatially) uncorrelated when the points are
further apart than about the distance~$\xi_2$, then the
earth's weather system might be understood as
consisting of weakly interacting, statistically
independent subsystems of size~$\xi_2$.  The number~$N$
of such subsystems can be estimated from the ratio of
the earth's surface area to to the area of an
approximately circular subsystem:
\begin{equation}
  N \approx \frac{4 \pi R_e^2}{\pi \xi_2^2}
    \approx \frac{4 \pi 6400^2}{\pi 1500^2}
    \approx 70
  , \label{n-of-weather-subsystems}
\end{equation}
where we have used the value~$R_e = 6400 \, \rm km$ for
the radius of the earth. Since each subsystem has its
own complex internal dynamics with many degrees of
freedom, \Eqref{n-of-weather-subsystems} would provide
a lower bound to the fractal dimension of the earth's
weather.  The value of~$70$ is consistent with the
failure of time-series-based algorithms
\cite{Abarbanel93} to estimate the fractal dimension of
the weather since these algorithms fail to converge for
fractal dimensions larger than about five
\cite{Pool89,Lorenz91}.

Similarly, a recent cardiology experiment has provided
some of the first spatiotemporal voltage measurements
on the surface of a pig's heart that was ingeniously
kept in sustained fibrillation for many minutes
\cite{Bayly93heart}. From this data, a two-point
correlation function was calculated which, although it
did not decay exponentially, did fall off sufficiently
rapidly so as to suggest a two-point correlation
length~$\xi_2 \approx 6 \, \rm mm$. From the average
radius of the pig heart~$R_{\rm pig} \approx 25 \, \rm
mm$, the arguments leading to
\Eqref{n-of-weather-subsystems} can be used to estimate
the number~$N$ of independent subsystems on the surface
of the fibrillating heart, which turns out to be~$N
\approx 70$. (The appearance of the number~$70$ in both
estimates is purely coincidental and has no known
religious significance.)  As a lower bound to the
fractal dimension of the heart, the value~70 is
consistent with the inability of researchers to detect
low-dimensional dynamics in ventricular fibrillation by
traditional time series methods \cite{Witkowski95}.

Is it reasonable to expect the two-point correlation
length~$\xi_2$ to provide dynamical information as
suggested above, which could cut the Gordian knot of
not being able to calculate large fractal dimensions
from data? A simple argument suggests that $\xi_2$
and~$D$ are generally independent
quantities\footnote{Another argument is that the
  fractal dimension is dynamically invariant while
  correlation functions and correlation lengths are
  not.}. One can disorder a phonograph record by
drilling holes into it in such a way so as to impose
any desired correlation length~$\xi_2$. By then
rotating the phonograph record rigidly with an angular
frequency~$\omega(t)$ derived from some $D$-dimensional
dynamical system, time series of any desired fractal
dimension can be created for any specified correlation
length~$\xi_2$.  \Figref{fig:convection-patterns} and
related time evolutions suggest, however, that such
phonograph dynamics is not representative of
spatiotemporal chaotic states which typically involve
the non-rigid motion of defects. A quantitative
exploration of a relation between the fractal
dimension~$D$ and length~$\xi_2$ might then be
insightful.

\subsection{The Dimension Correlation Length~$\xi_\delta$}
\label{xi-delta-calculations}

A possibly different length scale associated with
spatiotemporal chaos can be found by thinking about
attractors in phase space, rather than about the
disorder of fields such as~$u(t,x)$. As David Ruelle
first pointed out in~1982 \cite{Ruelle82}, widely
separated subsystems of a turbulent flow should be
weakly correlated in which case the spectrum of
Lyapunov exponents~$\lambda_i$\footnote{Recall that a
  bounded attractor of a dynamical system in a
  $N$-dimensional phase space has~$N$ real-valued
  Lyapunov exponents~$\lambda_i$ which are labeled in
  decreasing order~$\lambda_1 \ge \lambda_2 \ge \cdots
  \ge \lambda_N$ \cite{Ott93}. The sum of the first~$k$
  exponents, $\sum_{i=1}^k \lambda_i$, gives the
  average rate of exponential growth of the volume of
  an infinitesimal $k$-dimensional simplex of~$k+1$
  points. A system is dissipative if phase-space
  volumes contract on average, i.e., if the sum of
  all~$N$ Lyapunov exponents is negative. For
  infinite-dimensional phase spaces of dissipative
  partial differential equations, the dynamics often
  collapses to a finite $N$-dimensional subspace in
  which case these same ideas can be used.} for the
entire system should be the union of exponents
associated with each of the non-interacting subsystems.
This implies that the spectrum should be intensive in
the sense that~$\lambda_i = f(i/V)$ is a function only
of an intensive index~$i/V$ where~$V$ is the volume of
the system. This further implies (via
\Eqref{eq:dim-defn} below) that the fractal
dimension~$D$ of an attractor should be extensive, i.e,
$D \propto V$, for sufficiently large system size~$L$,
in which case we say that the dynamical system has {\em
  extensive chaos}.  The extensivity of~$D$ was first
demonstrated numerically by Paul Manneville in~1985 for
chaotic solutions of a one-dimensional partial
differential equation, the Kuramoto-Sivashinsky
equation \cite{Manneville85}.

Extensivity of~$D$ has since been established
numerically in other one- and two-dimensional
mathematical models (see \Figref{fig:extensive-chaos}
below) and in fact has been proposed by Cross and
Hohenberg to be the essential defining feature of
spatiotemporal chaos \cite[Page 945]{Cross93}. This
definition presently has the serious drawback of not
being computable with experimental data. Only one
numerical method is known for calculating fractal
dimensions of high-dimensional dynamical systems which
is the Kaplan-Yorke formula for the Lyapunov
dimension~$D_\mathcal{L}$ in terms of the spectrum of
Lyapunov exponents:
\begin{equation}
  D_\mathcal{L} = K
    +  \frac{1}{|\lambda_{K+1}|}  \sum_{i=1}^K \lambda_i
  , \label{eq:dim-defn}
\end{equation}
where the integer~$K$ is the largest integer such that
the sum of the first~$K$ Lyapunov exponents is
nonnegative\footnote{If we denote the partial sum of
  the first~$k$ exponents by $s_k = \sum_{i=1}^k
  \lambda_i$, then~$s_1 > 0$ by the assumption of chaos
  and~$s_N < 0$ by the assumption of dissipation. There
  is then some largest integer~$K$ for which~$s_K \ge
  0$ and~$s_{K+1} < 0$.  \Eqref{eq:dim-defn} is
  obtained by linearly interpolating to find the
  position of the root of the function~$s_k$.}. The
formula for~$D_\mathcal{L}$ has been proved to be equal
to the information dimension~$D_1$ in some cases
\cite{Ott93}, but it is not known whether this equality
holds for extensively chaotic systems.  With present
algorithms based on time series \cite{Abarbanel93},
only the first few largest exponents can be calculated
reliably, which usually are not enough to determine the
integer~$K$ in \Eqref{eq:dim-defn}. If dynamical
equations are known explicitly (so that the equations
can be linearized analytically around some given
nonlinear solution), then all the Lyapunov exponents
can be calculated numerically by studying directly the
exponential growth of hypervolumes of infinitesimal
simplices in the tangent space of some given nonlinear
orbit \cite{Parker89}. The computational effort to find
enough exponents to use \Eqref{eq:dim-defn} (all the
positive and enough negative ones) grows only
algebraically with the
dimension~$D_\mathcal{L}$\footnote{The exponent of this
  algebraic growth in computational effort is not known
  and is related to the issue of spatial ergodicity,
  whether one can trade off long integration times with
  short integration times in larger systems
  \cite{OHern96}.}, as opposed to
exponentially with~$D_\mathcal{L}$ for algorithms based
on time series. Even with this algebraic growth, most
calculations of~$D_\mathcal{L}$ are computationally
intense and only recently, with the increasing
availability of parallel computers, have researchers
been able to calculate the Lyapunov spectrum and
dimension of many one- and two-dimensional extensively
chaotic mathematical models.

Since the dimension~$D_\mathcal{L}$ grows in proportion
to the system volume~$V$ for large enough~$V$, the
dimension is itself not an interesting quantity since
it merely indicates the size of the system. More useful
is to define an intensive Lyapunov dimension
density~$\delta$:
\begin{equation}
  \delta = \lim_{V \to \infty} D_\mathcal{L} / V
  , \label{eq:delta-defn}
\end{equation}
which can be thought of as the number of active degrees
of freedom per unit volume. Since the dimension density
has physical units of inverse volume, Cross and
Hohenberg \cite{Cross93} suggest going one step further
by defining a dimension correlation
length~$\xi_\delta$:
\begin{equation}
  \xi_\delta = \delta^{-1/d} , \label{eq:xi-delta-defn}
\end{equation}
where~$d$ is the dimensionality of the system (e.g.,
$d=2$ for a large-aspect-ratio convection
experiment). This length can then be readily compared
with other lengths such as the critical
wavelength~$\lambda_c$, the two-point correlation
length~$\xi_2$, and the largest lateral system
size~$L$.

Two examples may help to interpret the meaning
of~$\xi_\delta$, which may be thought of as the radius
of a volume that contains one degree of freedom. If a
one-dimensional system of length~$L$ has a stable limit
cycle as the length~$L$ is varied (e.g., a stable plane
wave), then the dimension~$D=1$ (any limit cycle has
dimension one), the dimension density $\delta = 1/L$ is
a small number that vanishes with increasing~$L$,
and~$\xi_\delta = 1/\delta= L$ is a big length equal to
the size of the system. An opposite extreme would be a
one-dimensional coupled map lattice with~$L$ sites,
with each site containing a chaotic tent map of
dimension~$D=1$. In the limit of zero coupling, the
dimension~$D \to L$, $\delta \to 1$, and~$\xi_\delta
\to 1$ approaches the distance between neighboring
sites and is small compared to the system size~$L$. It
would be interesting to determine the values
of~$\xi_\delta$ for the two chaotic states of
\Figref{fig:convection-patterns}, e.g., from
simulations with two-dimensional Generalized
Swift-Hohenberg equations \cite{Xi95} or indirectly
from experimental data by using the extensive scaling
of the Karhunen-Lo\`eve decomposition
\cite{Zoldi96kld}.  A reasonable conjecture is that
\Figref{fig:convection-patterns}(b) will have a smaller
value of~$\xi_\delta$ than
\Figref{fig:convection-patterns}(a).

\subsection{A Comparison of Two-Point and Dimension Correlation Lengths}
\label{comparison-of-lengths}

Are the two correlation lengths ~$\xi_2$
and~$\xi_\delta$ related for homogeneous spatiotemporal
chaos, i.e., does average spatial disorder determine
dynamical complexity?  It is unlikely that they would
be equal to each other but a falsifiable hypothesis is
that they are proportional to one another, $\xi_2
\propto \xi_\delta$, in which case they should be
considered to be equivalent. One way to test this
hypothesis is to find a mathematical model of a
sustained nonequilibrium system for which one of these
quantities varies strongly with parameters and then to
compare one with the other. A hint of how to find such
a model comes from the theory of critical phenomena,
which tells us that the correlation length~$\xi_2$
diverges to infinity near a second-order phase
transition of an equilibrium system \cite[Section
148]{Landau80}. This observation poses several
difficult physics questions: what is meant by a phase
transition in an infinite nonequilibrium
system\footnote{For equilibrium systems, a phase
transition corresponds to a non-analytic behavior of
the free energy (as a function of thermodynamic
parameters like the temperature~$T$) that occurs {\em
only} in the thermodynamic limit of infinite system
size. Nonequilibrium systems do not have a free energy
and often have non-analytic behavior in order
parameters even in finite systems.}?  Do any examples
exist and, if so, do nonequilibrium phase transitions
have critical properties?

Two recent candidates for such models have been
recently studied by my group at Duke, the
one-dimensional complex Ginzburg-Landau (CGL) equation
\cite{Egolf94nature,Egolf95prl,Egolf94thesis} and the
two-dimensional Miller-Huse model
\cite{OHern96}. For simplicity, I briefly
review some results for the Miller-Huse model, which
more clearly succeed in falsifying the above
hypothesis, and which are also interesting in their own
right \cite{OHern96}.  The discussion will
also help to illustrate some of the ideas discussed in
previous sections such as how a system approaches
extensive behavior and how the length~$\xi_\delta$ is
calculated.

The Miller-Huse model is a clever deterministic
nonequilibrium version of the famous two-dimensional
Ising model \cite{Landau80}, which itself models the
continuous onset of ferromagnetic order for discrete
magnetic spins (i.e., when the spins all align in the
same direction), as the temperature is decreased from
the paramagnetic (unmagnetized) phase. The model is a
coupled map lattice (CML) in which space and time are
discretized and the lattice variables have continuous
real values. (There is no meaningful continuum limit of
this model as the lattice spacing and time step go to
zero, as would be the case for the numerical
discretization of a partial differential equation.)  On
a finite two-dimensional $L \times L$ square lattice
with periodic boundary conditions, a
piecewise-linear\footnote{Piecewise-linear chaotic maps
have uniform measures in phase space which makes them
useful for simplifying the mathematical analysis of a
chaotic system.} one-dimensional chaotic map~$y_{i+1} =
\phi(y_i)$ is placed on each of the $L^2$~lattice
sites. The output of each map is then coupled to its
nearest neighbor values by a simple linear diffusive
coupling with strength~$g$.  The model is not intended
to correspond closely to any physical system but, in
the spirit of the Ising model and of critical
phenomena, some details at long wavelengths are
expected to be independent of the lattice and of the
lattice map, and so might correspond to actual physical
systems.  That a phase transition might occur can be
understood as a competition between local chaos, which
generates disorder, and diffusion, which tends to damp
disorder.

\begin{figure}[ht] 
  \centering
  \includegraphics[width=3.5in]{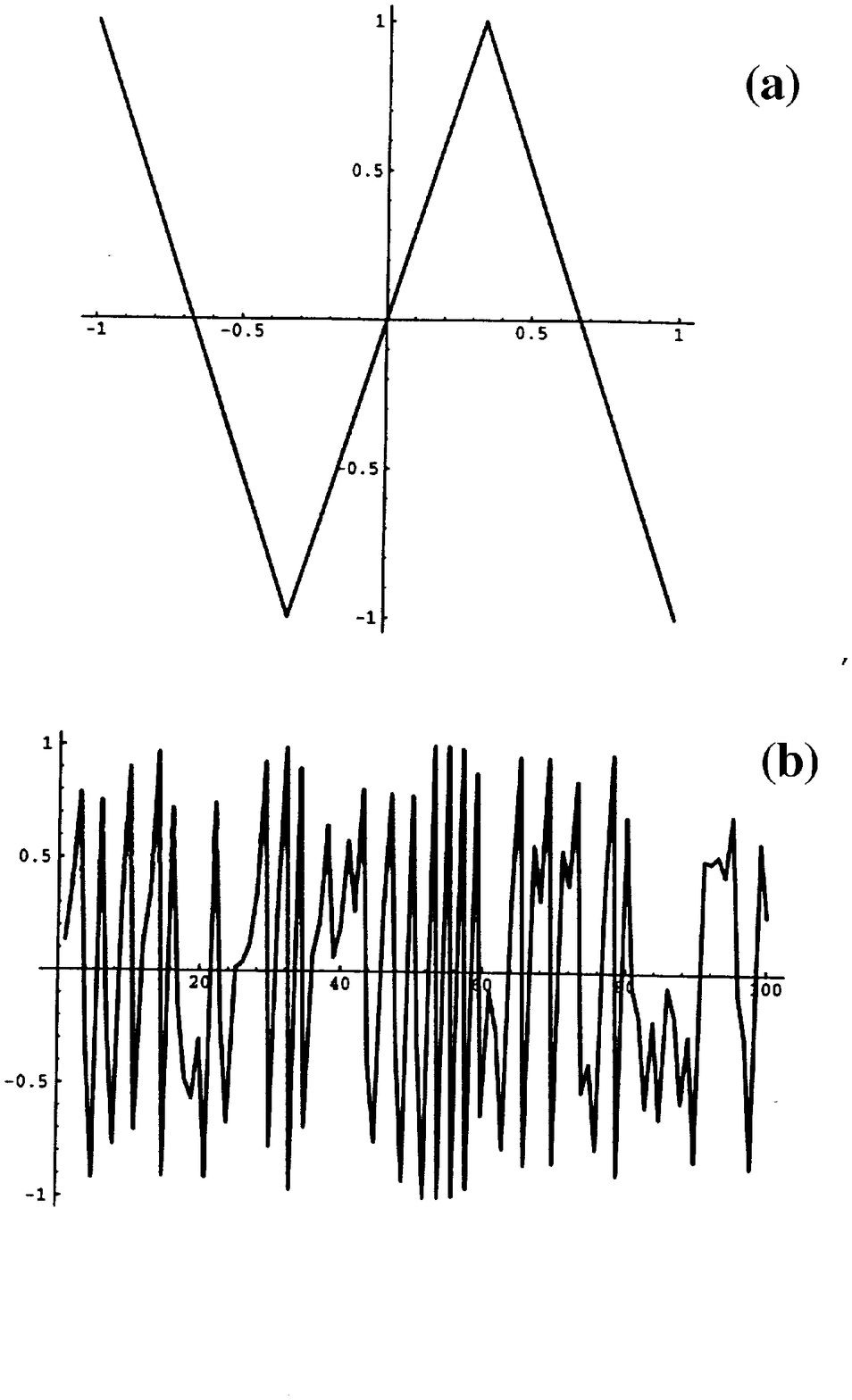}
\caption{ {\bf (a)} Plot of the odd Miller-Huse
  map~$y_{i+1} = \phi(y_i)$, \Eqref{eq:Miller-Huse-map}
  of the interval~$[-1,1]$ into itself. The map has a
  slope of constant magnitude~3 everywhere and is
  chaotic for nearly all initial conditions in the
  interval~$[-1,1]$.  {\bf (b)} Time series~$y_i$
  versus~$i$ for 100~iterations of the Miller-Huse map,
  illustrating its chaotic behavior. }
\label{fig:Miller-Huse-map}
\end{figure}

The CML is defined mathematically in the following way.
Let us denote by $y^t_i \in [-1,1]$ the variable at
spatial site~$i$ at integer time~$t$ (with $t=0$, $1$,
$\cdots$), and then define the following rule for
updating each lattice variable synchronously to
time~$t+1$ \cite{Miller93}:
\begin{equation}
  y_i^{t+1} = \phi( y_i^t )
    + g \sum_{j(i)} \left(
          \phi( y_j^t )
        - \phi( y_i^t )
      \right)
  ,  \label{eq:the-cml}
\end{equation}
where the parameter~$g$ is the spatial coupling
constant and where the sum goes over the four
indices~$j(i)$ that are nearest neighbors to site~$i$.
The function~$\phi$ defining the local map is given by
\begin{equation}
\phi(y) = \left\{
  \begin{array}{ccrcccr}
    -2 - 3y &  \mbox{for} &  \mbox{$-1$\ } & \le & y & \le & -1/3 ,\\
       3y   &  \mbox{for} & -1/3 & \le & y & \le &  1/3 ,\\
     2 - 3y &  \mbox{for} &  1/3 & \le & y & \le & \mbox{$1$,\ }
  \end{array}
  \right.
  \label{eq:Miller-Huse-map}
\end{equation}
and is shown in \Figref{fig:Miller-Huse-map}, together
with a representative time series of its chaotic
dynamics.
%%%%%%%%%%%%%%%%%%%%%%%%%%%%%%%%%%%%%%%%%%%%%%%%%%%%%%%%%%%%%%%%%
%%%%%%%%%%%%%%%%%%%%%%%%%%%%%%%%%%%%%%%%%%%%%%%%%%%%%%%%%%%%%%%%%
It maps the interval~$[-1,1]$ into itself and is
chaotic since the map has a slope of constant magnitude
equal to~3 which is everywhere greater than one. A
crucial insight of Miller and Huse was to choose the
local map \Eqref{eq:Miller-Huse-map} to have {\em odd}
parity, $\phi(-y) = - \phi(y)$, which turns out to be
necessary (but not sufficient) to allow an Ising-like
phase transition. Positive and negative values of
lattice variables then correspond to the up-spin and
down-spin magnetic variables of the Ising model.

\begin{figure}[ht] 
  \centering
  \includegraphics[width=4in]{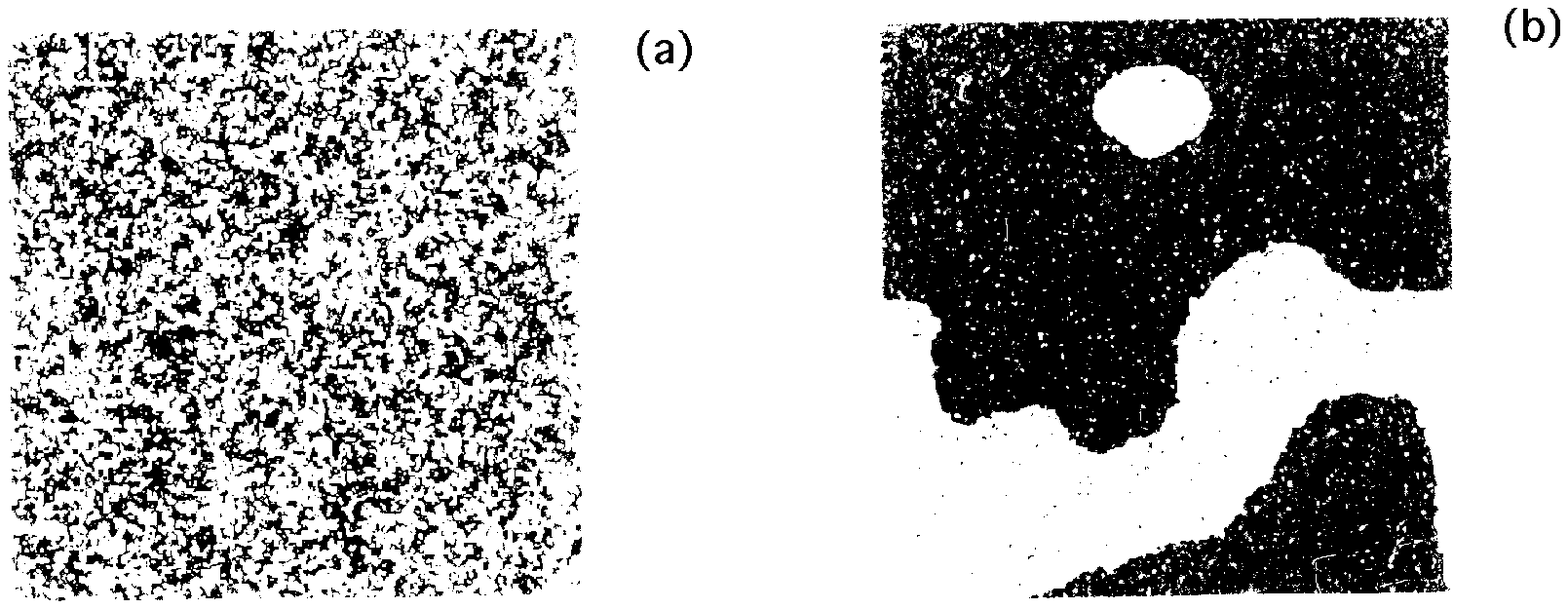}
\caption{ Onset of long-range spatial order in
  the two-dimensional Miller-Huse CML on a periodic
  $128 \times 128$ square lattice, as the coupling
  constant~$g$ is increased from below to above the
  phase transition point at~$g_c=0.205$.  Sites with
  positive and negative lattice values are colored red
  and green respectively. {\bf (a)} Instantaneous
  lattice values for $g=0.190$, for which the two-point
  correlation length~$\xi_2 = 3$ is comparable to the
  lattice spacing.  {\bf (b)} Instantaneous lattice
  values for~$g=0.235$ for which~$\xi_2$ is comparable
  to the size of the system. The two green droplets
  eventually annihilate leaving a homogeneous
  ``ferromagnetic'' red state. }
\label{fig:Miller-Huse-droplets}
\end{figure}

As a function of the lattice coupling parameter~$g$,
the qualitative behavior of this CML can be described
as follows. If one initializes each lattice
variable~$y^0_i$ inside the interval~$[-1,1]$ and
provided~$g$ is smaller than~$1/4$, the dynamics will
be bounded forever. For small values of~$g$, the
dynamics settle down after some transient into a
spatiotemporal chaotic state consisting visually of
small domains of sites that all have negative or
positive values together; these regions of similar sign
are called ``droplets'' by analogy to the clusters of
up- and down-spins found in the Ising model
(\Figref{fig:Miller-Huse-droplets}).
%%%%%%%%%%%%%%%%%%%%%%%%%%%%%%%%%%%%%%%%%%%%%%%%%%%%%%%%%%%%%%%%%
%%%%%%%%%%%%%%%%%%%%%%%%%%%%%%%%%%%%%%%%%%%%%%%%%%%%%%%%%%%%%%%%%
These small fluctuating droplets corresponds to a
high-temperature disordered paramagnetic phase. As~$g$
is increased towards the critical value\footnote{If one
  defines an average lattice magnetization $M =
  \langle{\rm sign(y^t_i)} \rangle$ as the space-time
  averaged sign of the lattice variables, then the
  critical value~$g_c$ can be accurately determined by
  studying the critical scaling~$(g - g_c)^\alpha$ of
  the magnetization~$M$ as it bifurcates from zero (the
  disordered paramagnetic phase) to a finite positive
  or negative value (onset of the ordered ferromagnetic
  phase) \cite{Miller93,Marcq96}.}
$g_c=0.205$, the dynamics remain chaotic and the
droplets grow in size.  Correspondingly, the
correlation length~$\xi_2$ of the field values~$y^t_i$
starts to increase until it becomes of order the system
size at the critical value~$g_c$.  For still larger
values of~$g$, one finds just one or two droplets
spanning the entire lattice, corresponding to a
low-temperature ferromagnetic, ordered phase.  The idea
is then to compare the lengths~$\xi_2$ and~$\xi_\delta$
across the nonequilibrium transition point of~$g=0.205$
as~$\xi_2$ diverges to infinity.

\begin{figure}[ht] 
  \centering
  \includegraphics[width=3.5in]{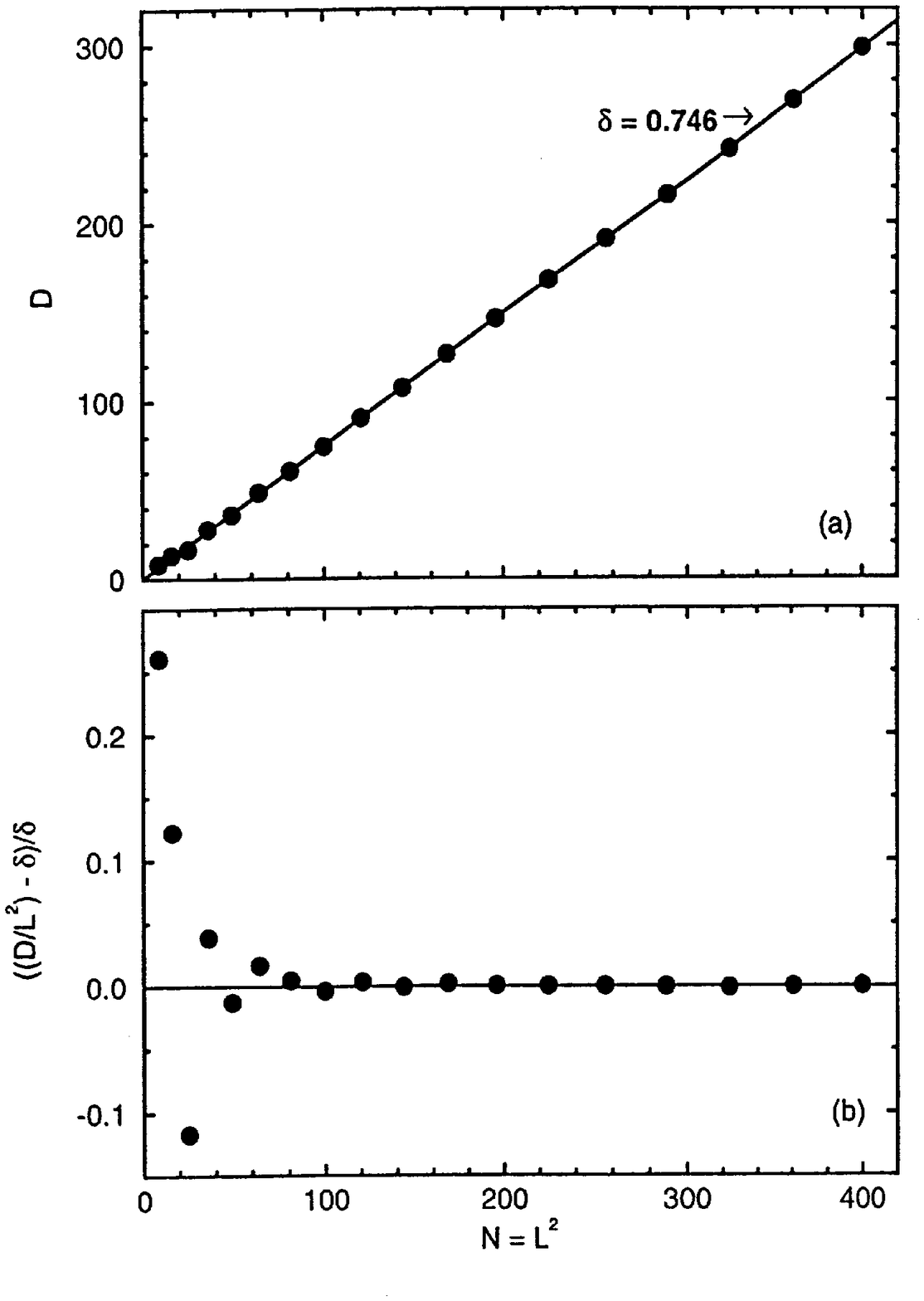}
\caption{ Lyapunov fractal dimension~$D$ versus the
area of the lattice~$N=L^2$ for the periodic
Miller-Huse CML \Eqref{eq:Miller-Huse-map} and for
coupling constant~$g=0.199$. The dimension~$D$ was
calculated from the Kaplan-Yorke formula
\Eqref{eq:dim-defn} by iterating the CML for
~$T=50,000$ iterations after transients died out.  {\bf
(a)} The dimension grows linearly and extensively with
lattice area for~$N > 100$. The slope of this curve for
large~$N$ gives a dimension density~$\delta = 0.746$,
corresponding to a dimension correlation
length~$\xi_\delta = \delta^{-1/2} = 1.16$.  {\bf (b)}
The normalized deviation $(D/N - \delta)/\delta$, of
the dimension density~$\delta$ from the value~$D/N$,
illustrates the rapid approach to extensive scaling for
small systems sizes~$L > 9$.  }
\label{fig:extensive-chaos}
\end{figure}

For coupling constant~$g=0.199$,
\Figref{fig:extensive-chaos} shows that the Miller-Huse
CML is, in fact, extensively chaotic as the lattice
area (total number of lattice sites~$N$) is increased.
%%%%%%%%%%%%%%%%%%%%%%%%%%%%%%%%%%%%%%%%%%%%%%%%%%%%%%%%%%%%%%%%%
%%%%%%%%%%%%%%%%%%%%%%%%%%%%%%%%%%%%%%%%%%%%%%%%%%%%%%%%%%%%%%%%%
For small~$N$, the Lyapunov dimension~$D_\mathcal{L}$
behaves irregularly, which corresponds to the sensitive
dependence of low-dimensional dynamics on parameter
changes. For lattices sizes~$L$ larger than about~9,
the extensive regime begins and $D_\mathcal{L}$ grows
linearly with~$N$. The dimension density~$\delta =
0.746$ is obtained from a least-squares fit of the
linear part of the curve and this value corresponds to
a dimension correlation length of~$\xi_\delta =
\delta^{-1/d} = \delta^{-1/2} = 1.16$ which is
comparable to one lattice spacing.
\Figref{fig:extensive-chaos}(b) shows that the onset of
extensive behavior is not abrupt. The deviation from
extensivity decays rapidly and non-monotonically.

\begin{figure}[ht] 
  \centering
  \includegraphics[width=3.5in]{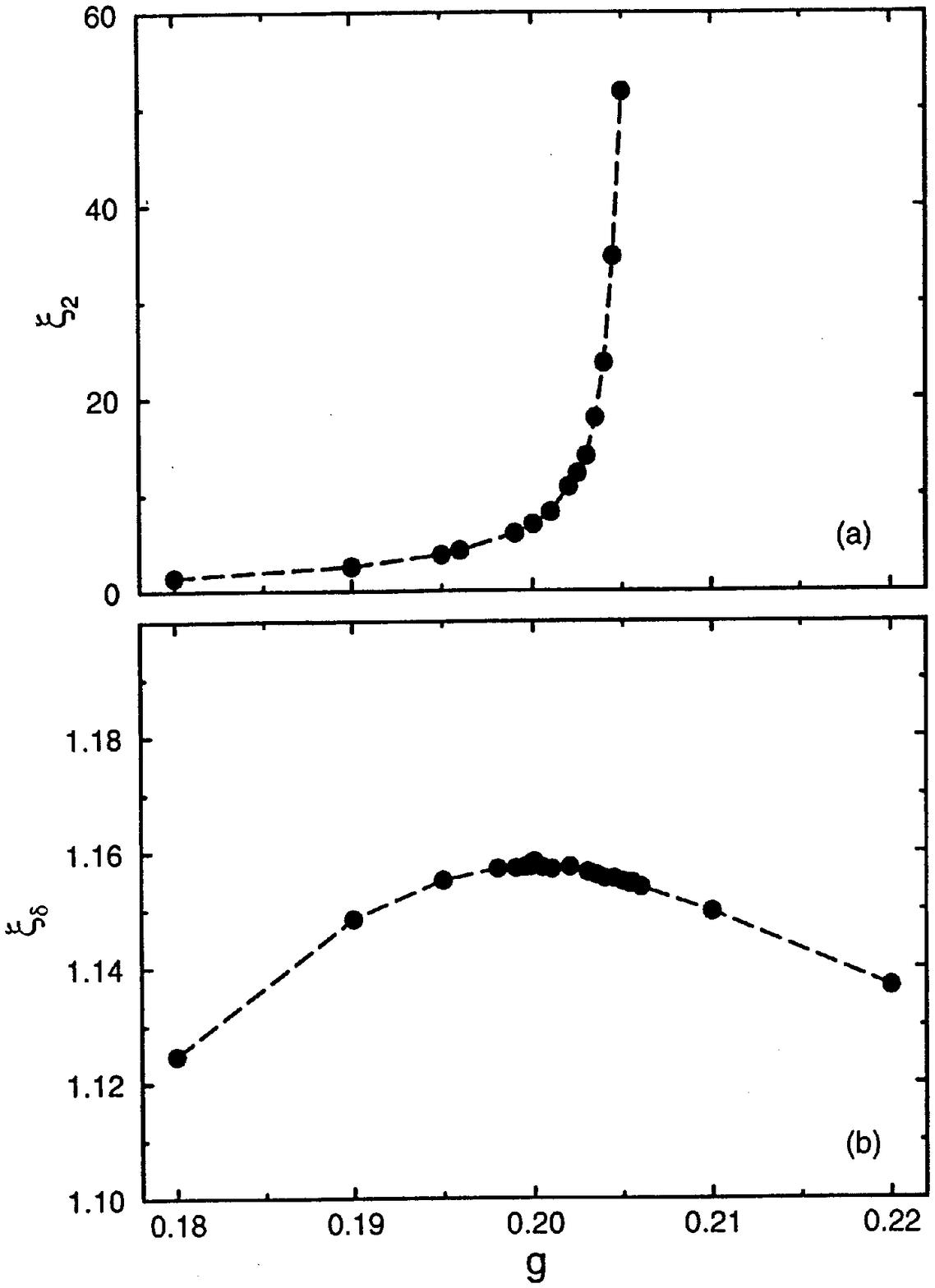}
\caption{ Variation of {\bf (a)} the two-point
correlation length~$\xi_2$ and {\bf (b)} the dimension
correlation length~$\xi_\delta$ with the coupling
constant~$g$ for the Miller-Huse CML on a square
periodic lattice. The length~$\xi_2 \propto (g -
g_c)^{-\nu}$ diverges with an exponent~$\nu \approx 1$
\cite{Marcq96} near the critical value~$g_c=0.205$
while the length~$\xi_\delta$ is bounded and varies
smoothly, attaining a local maximum at a
value~$g=0.200$ that is distinctly less than~$g_c$.
Lattice sizes of up to~$L=1024$ were used to calculate
these lengths, with integration times of up to~$T=3
\times 10^5$. }
\label{fig:xi_2-vs-xi_delta}
\end{figure}

\Figref{fig:xi_2-vs-xi_delta} is the most important
figure in this paper and shows how the two lengths,
$\xi_2$ and~$\xi_\delta$, vary in the vicinity of the
Miller-Huse nonequilibrium phase transition
at~$g_c=0.205$.
%%%%%%%%%%%%%%%%%%%%%%%%%%%%%%%%%%%%%%%%%%%%%%%%%%%%%%%%%%%%%%%%%
%%%%%%%%%%%%%%%%%%%%%%%%%%%%%%%%%%%%%%%%%%%%%%%%%%%%%%%%%%%%%%%%%
As is the case with the Ising model, the two-point
correlation length diverges at the critical
value~$g_c$, $\xi_2 \propto (g - g_c)^{-\nu}$, with
critical exponent~$\nu \approx 1$ in accord with the
arguments of Miller and Huse \cite{Marcq96}. In
contrast, \Figref{fig:xi_2-vs-xi_delta}(b) shows
that~$\xi_\delta$ over the same range of~$g$ is bounded
and varies smoothly over the small range of~1.12
to~1.16, attaining a local maximum\footnote{The
proximity of this maximum to~$g_c$ is a coincidence for
the two-dimensional square lattice.  Other calculations
of the Miller-Huse model for two-dimensional hexagonal
and for three-dimensional cubic lattices show that the
local maximum in~$\xi_\delta$ is unrelated to~$g_c$
\cite{OHern96}.} at the value~$g=0.200$ which
is distinctly less than the value~$g_c$ at which the
transition occurs.

Despite the dramatic ordering of lattice variables with
increasing~$g$ as depicted in
\Figref{fig:Miller-Huse-droplets}, we conclude that
there is no relation between the average spatial
disorder (as measured by the diverging two-point
correlation length~$\xi_2$) and the dynamical
complexity of the corresponding strange attractor (as
measured by the intensive dimension correlation
length~$\xi_\delta$).  At least two length scales, and
likely an infinity of lengths scales, are needed to
characterize homogeneous spatiotemporal chaos. An
unfortunate corollary is that the length~$\xi_2$ can
{\em not} be used to estimate the dimension
density~$\xi_\delta$ and hence the fractal dimensions
of complex dynamical systems like the weather and the
heart, as originally suggested by the heuristic
arguments associated with
\eqref{n-of-weather-subsystems}.

Other calculations confirm these conclusions. Thus
calculations for different system sizes~$L$ of a
one-dimensional partial differential equation, the
complex Ginzburg-Landau (CGL) equation, show similarly
that over a certain parameter range, the length~$\xi_2$
can grow rapidly to large values while the
length~$\xi_\delta$ remains moderate and nearly
constant \cite{Egolf95prl,Egolf94thesis}.  Calculations
on a certain CML \cite{Bhagavatula92} have shown that
systems with generic algebraic decay of spatial
correlations, for which the length~$\xi_2$ is
effectively infinite over a continuous range of
parameters, can still be extensively chaotic, with a
finite dimension correlation length~$\xi_\delta$
\cite{OHern97}. An interesting implication of
these recent calculations is that~$\xi_\delta$ always
seems of moderate size, often comparable to the
critical wavelength~$\lambda_c$ of a pattern-forming
instability and small compared to the two-point
correlation length~$\xi_\delta$. This suggests that
many large-aspect-ratio experiments such as those shown
in \Figref{fig:convection-patterns} are already in the
extensively chaotic regime and can be used to test
future theoretical work on extensive chaos.

These results present two interesting theoretical
puzzles. One is to discover what mathematics and
physics determines the magnitude of the
length~$\xi_\delta$. For the Miller-Huse CML, the
value~$\xi_\delta \approx 1$ says that the dynamics of
each lattice map~$\phi(y)$ is effectively independent
of all others, even though the diffusive coupling leads
to the onset of long range order as measured
by~$\xi_2$. This is somewhat plausible since there is a
strong source of local chaos even in the absence of
coupling between neighbors. The more mysterious
situation is for systems with continuous spatial
variables such as the CGL~equation and for the
convecting fluid in \Figref{fig:convection-patterns}.
In these cases, there is no chaos except in the
presence of spatial coupling; sufficiently small
isolated subsystems have either stationary or periodic
nontransient dynamics.

The second theoretical puzzle is whether there is some
way to estimate the dimension correlation
length~$\xi_\delta$ directly from finite amounts of
imprecise experimental data. The calculations described
above for the Miller-Huse CML used a brute-force
computationally-intense method that first required
calculating the Lyapunov fractal dimension of the
entire system from known dynamical equations for
several different volumes, and then extracting the
length~$\xi_\delta$ from the extensive scaling. This is
utterly impractical for spatiotemporal data of the sort
represented in \Figref{fig:convection-patterns} since
there is no known way of calculating many Lyapunov
exponents from time series recorded from
high-dimensional systems.  Instead, one might hope to
exploit the fact that~$\xi_\delta$ is an intensive
quantity and so might depend only on information
localized to some region of space.

There have been two recent attempts to estimate the
value of~$\xi_\delta$ from time series by
generalizations of the Procaccia-Grassberger algorithm
\cite{Ott93} for low-dimensional systems
\cite{Tsimring93,Bauer93} but these efforts lack a
rigorous justification and their numerical convergence
is not convincing.  More recently, Scott Zoldi and the
author have proposed using a variation of the
Karhunen-L\`oeve decomposition (KLD)
\cite{Preisendorfer88}, in which the KLD is applied to
data in concentric volumes of increasing radius and
then the KLD eigenvalues from these different
calculations are combined to provide an estimate of an
intensive correlation length~$\xi_{\rm KLD}$ which has
properties similar to the dimension correlation
length~$\xi_\delta$ \cite{Zoldi96kld}. While this
approach can not quantitatively predict the value
of~$\xi_\delta$, calculations for the Miller-Huse model
show that the length~$\xi_{\rm KLD}$ behaves
qualitatively like~$\xi_\delta$ as the coupling~$g$ is
varied, and so might serve the purpose of an easily
calculated, localized order parameter for
characterizing large, weakly inhomogeneous,
spatiotemporal chaotic systems.

\section{Conclusions}
\label{conclusions}

This survey article has discussed recent and ongoing
efforts to understand how one particular
mechanism---increase of system size---affects the
complexity of nontransient homogeneous sustained
nonequilibrium systems.  Perhaps the single most
important idea from this discussion is that of {\em
  extensive chaos}: for sufficiently large homogeneous
chaotic systems, the fractal dimension~$D$ (effective
number of degrees of freedom) becomes proportional to
the volume~$V$ of the system as illustrated in
\Figref{fig:extensive-chaos}. This mathematical
statement can be interpreted physically as saying that
big homogeneous systems become complex in a simple way
with increasing size, by replicating weakly-interacting
and statistically similar subsystems of some
characteristic size that we have defined as the
dimension correlation length~$\xi_\delta$. Another
interpretation of extensive chaos is that there are no
interesting collective effects with growing size, as
might be expected if small dynamical units somehow bind
themselves into a larger effective unit with fewer
total degrees of freedom. Collective effects would
possibly appear as a proportionality~$D \propto
L^\alpha$ (where $L$ is the system size), with an
exponent~$\alpha$ that is not the dimensionality~$d$ of
the system.

Extensive chaos is important because it shifts
attention away from characterizing a system by global
quantities (e.g., by its fractal dimension), towards
understanding the properties of a subsystem and how
different subsystems interact.  Research along these
lines for sustained nonequilibrium systems is still
only in its infancy
\cite{Zaleski89,Bourzutschky92,Miller93,Egolf95prl,Chow95,OHern96,Dankowicz96}
but should blossom in the near future if for no other
reason that computer power continues to increase
steadily, which will allow a wider variety of physical
problems to be simulated and studied mathematically,
and which will permit further quantitative comparisons
between theory and experiment.

Besides surveying recent progress in theory and
experiment, this review has also tried to point out
interesting open questions. I conclude this paper by
summarizing some of these questions, with the hope of
stimulating further research:
\begin{enumerate}

\item A theme throughout this paper was that big
homogeneous chaotic systems can be considered as a gas
of weakly interacting subsystems. It would be
insightful to find some way to demonstrate this idea
directly, rather than indirectly by the extensive
scaling of the fractal dimension. One idea would be to
study how localized short-lived infinitesimal
perturbations propagate in spatiotemporal chaotic
systems. Do they die off substantially on the length
scale of~$\xi_\delta$?

\item Can the idea of a subsystem provide a way to
  quantify the dynamics of inhomogeneous sustained
  nonequilibrium systems, at least for the simplified
  case of parameter values that are slowly varying in
  space? If a spatially dependent parameter~$p(x)$
  varies on a length scale large compared
  to~$\xi_\delta$, then one can speculate that a
  subsystem centered at position~$x_0$ would, over a
  certain range of length and time scales, be similar
  to a subsystem of an infinite homogeneous dynamical
  system whose spatially uniform parameter value
  corresponds to the value~$p_0=p(x_0)$. This idea was
  recently confirmed for a Miller-Huse model with a
  spatially varying coupling constant~$g$, using a
  correlation length based on the extensivity of the
  Karhunen-L\`oeve decomposition \cite{Zoldi96kld}.

\item The Navier-Stokes equations can be considered as
arising from a long-wavelength, long-time-scale
averaging of molecular
dynamics. \Figref{fig:convection-patterns}(b) suggests
an analogous reduction: over length scales large
compared to the dimension correlation
length~$\xi_\delta$ (the size of the spirals in this
figure?), can the intricate local dynamics be replaced
by a simple hydrodynamic-like equation, perhaps with a
Gaussian uncorrelated noise source arising from the
local chaos?  This has been successfully done in two
cases, for the one-dimensional Kuramoto-Sivashinsky
equation \cite{Chow95} and for a one-dimensional CML
with a conserved quantity
\cite{Bourzutschky92}. However, there are no principles
known for deriving such equations or even for
determining whether such equations exist.

\item Are intensive quantities like~$\xi_\delta$ {\em
    local} quantities that depend only on information
  in a finite region of space? If so, are there
  time-series based algorithms that can estimate these
  intensive quantities, without first calculating an
  extensive quantity?  The calculations of
  Ref.~\cite{Zoldi96kld} suggests that this should be
  possible but more work is needed.

\item What determines the time scale for transients to
  decay in sustained nonequilibrium systems? In
  particular, what is the mathematical origin of
  supertransients \cite{Crutchfield88,Lai95prl} for
  which the decay time grows exponentially with system
  size? A resolution of this question will be important
  for establishing the concept of subsystems in
  nonequilibrium systems, since subsystems should
  become nontransient over any experimental or
  computational observation time.

\item Can one use the idea of spatial ergodicity to
replace difficult-to-parallelize long-time integrations
by easier-to-parallelize short-term integrations in
large spatial systems? Especially interesting would be
a quantitative comparison of statistics obtained from a
simulation involving~$N$ independent realizations
(integrations of separate systems on separate
processors starting from different initial conditions)
with statistics obtained from one large system that
contains about~$N$ weakly interacting subsystems of
size~$\xi_\delta$.

\item Finally, will the concept of weakly interacting
  subsystems help to solve the problem of how to
  control large spatiotemporal chaotic systems by weak
  perturbations \cite{Hu95}, e.g., by stabilizing an
  unstable periodic orbit? Numerical experiments have
  shown that weak perturbations applied at a point can
  only control dynamics out to some radius from that
  point \cite{Aranson94}, with the implication that
  spatially distributed control sites are needed. It
  would be interesting to study whether the
  length~$\xi_\delta$ is a lower bound for the scale at
  which different control points should be separated in
  space, so that as few control points as possible are
  used.
\end{enumerate}

\section*{Acknowledgments}
The author would like to thank Dr. Katie Coughlin for
the invitation to attend this stimulating workshop. The
author would also like to thank Philip Bayly, Hugues
Chat\'e, David Egolf, Andrew Krystal, Corey O'Hern,
Matt Strain, and Scott Zoldi for stimulating and
insightful conversations.

%\bibliographystyle{amsalpha}
%\bibliography{biology,cardiology,cfd,chaos,control,cs,hsg,lasers,math,meteorology,physics,spatiotemporal,time-series}

\newcommand{\etalchar}[1]{$^{#1}$}
\providecommand{\bysame}{\leavevmode\hbox to3em{\hrulefill}\thinspace}

\end{document}